# 2+1 dimensional Fermions on the low-buckled honey-comb structured lattice plane and classical Casimir-Polder force

Partha Goswami*

Deshbandhu College, University of Delhi, Kalkaji, New Delhi-110019, India

*physicsgoswami@gmail.com

**Abstract**  We start with the well-known expression for the vacuum polarization and suitably modify it for 2+1 dimensional spin-orbit coupled (SOC) fermions on the low-buckled honey-comb structured lattice plane described by the low-energy Liu-Yao-Feng-Ezawa (LYFE)Model Hamiltonian involving the Dirac matrices in the chiral representation obeying the Clifford algebra. The silicene and germanene fit this description suitably. They have the Dirac cones similar to those of graphene and SOC is much stronger. The system could be normal or ferro-magnetic in nature. The silicene turns into the latter type if there is exchange field arising due to the proximity coupling to a ferro-magnet such as depositing Fe atoms to the silicene surface. For the silicene, we find that the many-body effects considerably change the bare Coulomb potential by way of the dependence of the Coulomb propagator on the real-spin, iso-spin, and the potential due to an electric field applied perpendicular to the silicene plane. The computation aspect of the Casimir-Polder force needs to be investigated in this paper. An important quantity in this process is the dielectric response function (DRF)of the material. The plasmon branch was obtained by finding the zeros of DRF in the long wave-length. limit. This leads to the plasmon frequencies. We find that the collective charge excitations at zero doping, i.e., intrinsic plasmons, in this system, are absent in the Dirac limit. The valley-spin-split intrinsic plasmons, however, come into being in the case of the massive Dirac particles with characteristic frequency close to 10 THz. Our scheme to calculate the Casimir-Polder interaction (CPI) of a micro-particle with a sheet involves replacing the dielectric constant of the sample in the CPI expression obtained on the basis of the Lifshitz theory by the static DRF obtained using the expressions for the polarization function we started with. Though the approach replaces a macroscopic constant by a microscopic quantity, it has the distinct advantage of the many-body effect inclusion seamlessly. We find the result that the for non-trivial susceptibility and polarizability values of the sheet and micro-particle , respectively, there is crossover between attractive and repulsive behavior . The transition depends only on these response functions apart from the ratio of the film thickness and the micro-particle separation (D/d) and temperature. Furthermore, there is a longitudinal electric field induced topological insulator(TI) to spin-valley polarized metal (SVPM) transition in silicene, which is also referred to as the topological phase transition (TPT). The low-energy SVP carriers at TPT possess gap-less (mass-less) and gapped (massive) energy spectra close to the two nodal points in the Brillouin zone with maximum spin-polarization. We find that the magnitude of the Casimir-Polder force at a given ratio of the film thickness and the separation between the micro-particle and the film is greater at TPT than at the topological insulator and trivial insulator phases.

## 1. Introduction

The purpose of this communication is to report the theoretical investigation of the Casimir-Polder interaction (CPI) (a variant of the Casimir interaction(CI))[1,2,3,4,5] of a micro-particle with a silicene sheet using the expressions for the polarization tensor and Coulomb propagator derived by the previous workers [6,7,8,9,10]. The zero and the finite temperature CPI/CI, and the ubiquitous, Van der Waals interactions[11,12,13] – the molecular counterpart of CI, are the important examples of the long-ranged, dispersive fluctuations induced interactions. Our strategy is to deploy this particle-sheet interaction for probing the out-of-plane electric field ($E_z$) induced phase transition in silicene from a topological insulator(TI) to a band insulator(BI)/trivial insulator with the spin-valley-polarized metal (SVPM) at a critical value of $E_z = E_{zc}$. For this purpose, we invoke the well-known [6] expression for the dynamic polarization function followed by the dielectric response function (DRF) properties. We examine the polarization function and the corresponding plasmon frequencies of the 2+1 dimensional spin-orbit coupled (SOC) fermions in the non-relativistic approximation on the low-buckled honey-comb structured lattice plane [7,8] as in silicene and germanene. The plasmon branch can be obtained by finding the zeros of the dielectric function. We then calculate the classical Casimir-Polder interaction (CPI) of a micro-particle with a silicene sheet using the DRF properties. The central point of the scheme of calculation involves replacing the (macroscopic) dielectric constant of the Si/Ge sample, in the CPI expression obtained on the basis of the Klimchitskaya-Mostepanenko theory [9,10, 14], by the (microscopic) static DRF obtained using the expressions for the polarization function we started with. Though the approach replaces a macroscopic constant by a microscopic quantity, it has the distinct advantage of the many-body effect inclusion seamlessly. The closed-form, precise expressions for the Klimchitskaya-Mostepanenko reflection coefficients of the electro-magnetic fluctuations on the silicene sheet plus substrate, for two independent modes, viz. the transverse magnetic (TM) and the transverse electric (TE) polarizations, involved in the CPI expressions here are given by the Fresnel coefficients [9,10,14,15] corresponding to the reflection on the boundary planes between the vacuum and the film material and also between the film material and the substrate. These Fresnel coefficients, on the other hand, are also given by the dielectric permittivity and the magnetic permeability, of the film and the substrate. In a more accurate and detailed analysis, within the framework of the Dirac model, the reflection coefficients of the electromagnetic oscillations from silicene needs to be expressed via the polarization tensor[9,10,14] of silicene in (2+1)-dimensions. At relatively large separation, the Casimir and Casimir-Polder free energy and force are then given by the zero-frequency contributions to respective Lifshitz formulas [16,17,18].

The silicene is basically a semimetal because the valence and conduction bands touch at the Fermi level. It has a larger spin-orbit coupling (SOC)induced gap than graphene [19,20,21]. It exhibits a tunable band gap due to an applied, out-of-plane electric field $E_z$. Remarkably, the tuning of $E_z$ is reflected in the change of state : the system makes a crossover from TI ($E_z < E_c$) to BI ($E_z > E_c$) with the SVPM state ($E_z = E_c$) in between. In fact, at ($E_z/E_c$) = 1.00, the gap-closing occurs in one of the spin-specified valence-conduction band pairs of one of the valleys(say, the one with the valley index $\xi = +1$) while the gap-opening occurs for the opposite spin band pair of the same valley. At the opposite valley ( with the valley index $\xi = -1$), there would be different scenario: the gap-closing (opening) for the opposite(same) spin band pair. Since the gap-closing occurs at ($E_z/E_c$) = 1.00, here-in-after we shall refer to this *TI-VSPM-BI* transition as the topological phase transition(TPT). The detection of the finger-print of TPT by employing the Friedel oscillation and collective excitation in silicene had been reported by Tabert et al.[22] and Chang et al[23]. The thrust area of these authors was the calculation of the dynamical polarization function and the undamped plasmon mode emerging from the single-particle excitation spectrum. Our aim is to present here an alternative way of finding signature of TPT in a measurable quantity, viz. in the Casimir-Polder interaction of the micro-particle with a silicene sheet using the expression for the polarization tensor obtained[6,7,8,9]. Since the Casimir effect depends on the electromagnetic response of a material, the tunability of the response ensured by change in the material properties implies that the casimir-effect on the material gets modified accordingly. Conversely, the casimir-effect modification will be an indicator of the change in the material properties. In this communication, our aim is to exploit precisely this idea to ascertain the onset of the topological phase transition(TPT).We expect that when the TPT takes place the Casimir-Polder free energy/interaction between an electrically and magnetically polarizable micro-particle with a silicene sheet will yield some noticeable change. Furthermore, in ref.[24] it has been shown that,(i) in the quantum Hall regime, the Casimir force asymptote for the large-separation is quantized, and (ii), remarkably, for the ambipolar systems, such as graphene, the force is electrically tunable between attractive and repulsive values. We shall discuss in a future communication how the latter is possible for the silicene/germanene as these systems are also ambipolar.

The only crystalline allotrope of silicon(Si) is perhaps a single layer of Si-atoms known as silicene. It exhibits an analogous honeycomb structure [25,26,27,28] as graphene. Recen-tly, it is also observed from first principles density functional theory calculations that ultra-thin nanosheets or single sheet of zinc oxide also exist in a graphene form with many interesting properties [29]. In addition, it is also experimentally observed that platinum intercalation at high temperature on silicon carbide leads to the formation of platinum silicide at the interface with similarities to graphene **[30]**. On the other hand, silicene has been synthesized on the Ag(111) substrate a few years ago. In a two-dimensional germanium(Ge) layer grown recently, by the deposition of Ge onto the Au(111) surface [**31**] similar to that of Si on Ag(111), clear emergence of the honeycomb structure has been noticed in one of the several possible phases. These discoveries have promoted vigorously the search for materials that may possess topologically insulating phase **[32,33,34]**. For the silicene and the germanene systems, the linear band crossing occurs at the *K($4\pi/3a$, 0)* and *K′* *(−$4\pi/3a$, 0)* symmetry points at opposite corners of the hexagonal Brillouin zone. The charge carriers consequently behave like relativistic particles with a conical energy spectrum and Fermi velocity $v_F \approx$ $1\times10^6$ m-s$^{-1}$ and $5\times10^5$ m/s, respectively, as in graphene. Yet another point of similarity is that the unit cell of the Si/Ge systems contain two atoms which gives rise to two different sub-lattices A and B as in graphene **[19,20,21]**.The states near the Fermi energy are $\pi$ orbitals residing near the Dirac points *K* and *K′* **[19,20,21]**. The honeycomb lattice of the system is distorted due to a larger (than carbon) ionic radius of silicon atom and forms a buckled structure. The *A* and *B* sites per unit cell form two sub-lattices separated by a perpendicular distance, say, *2ℓ*. The structure generates a staggered sub-lattice potential $\Delta_z = 2\ell E_z$ between silicon atoms at *A* sites and *B* sites for the applied electric field $E_z$ perpendicular to the system-plane. Additionally, we find that the quantum anomalous Hall state could be accessed in silicene by introducing an exchange field. The exchange field *′M'* arises due to proximity coupling to a ferro-magnet such as depositing Fe atoms to the silicene surface or depositing silicene to a ferromagnetic insulating substrate. We investigate here the Casimir-Polder interaction (CPI) of the micro-particle with a normal silicene ( $\Delta_z = 0$ and *M = 0*) sheet and a ferro-magnetic silicene ($\Delta_z \neq 0$, and $M \neq 0$) sheet using the expression for the polarization tensor[**1-9**] at zero frequency. The situation in detail is shown in Figure 1. We shall outline the scheme of the calculation of CPI in section3.

The mutual electromagnetic correlations between two spatially separated systems gives rise to Casimir/ Casmir-Polder effect. The corresponding forces, which are generally attractive for most vacuum-separated metallic or dielectric geometries, are due to the contribution to the ground-state [18,35] energy of the coupled system. It may be mentioned that the repulsive Casimir forces may also occur in plethora of systems, such as the fluid-separated dielectrics [36], composite metamaterials[37], and so on, as suggested theoretically. Experimentally, the forces have been realized for the first time involving test bodies immersed in a liquid medium- ethanol[36]. We investigate here the Casimir-Polder free energy corresponding to interactions of an electrically and magnetically polarizable micro-particle with a magneto-dielectric sheet and look for the repulsive Casimir-Polder forces between these dielectric materials possessing non trivial magnetic susceptibility and polarizabilities. We show that for such response function values, the crossover between attractive and repulsive behavior depends only on the sheet-impedance and micro-particle polarizability ratio apart from the ratio of the film thickness and the micro-particle separation (D/d) and temperature. For example, with these values ( $\varepsilon_0^{(1)}$ = 14, $Z_1$ = 0.5 , and $Z_2$ = (01, 02, 03, 04) ), we find that interaction is attractive as long as $\left(\frac{D}{d}\right) \lesssim 0.2$, or, $d \gtrsim 5\,D$. For $\left(\frac{D}{d}\right) > 0.2$ (or, $d < 5\,D$), the interaction is repulsive, while with the values ( $\varepsilon_0^{(1)}$ = 14, $Z_1$ = 1.0 , and $Z_2$ = (01, 02, 03, 04) ), we find that interaction is attractive as long as $\left(\frac{D}{d}\right) \lesssim 0.1$, or, $d \gtrsim 10D$. For $\left(\frac{D}{d}\right) > 0.1$ (or, $d < 10\,D$), the interaction is repulsive.

The paper is organized as follows: In Sec. II, the Casimir-Polder free energy corresponding to interactions of an electrically and magnetically polarizable micro-particle with a magneto-dielectric sheet is written down which contains relevant presentation of some features of the theory of Casimir-Polder interaction. A brief outline of the Liu-Yao-Feng-Ezawa(LYFE)model [19,20,21] of Si/Ge system is also given in this sub-section and the low-energy excitation spectrum is obtained. In Sec.III, the dynamical polarization function for the buckled honeycomb lattices in the intrinsic limit (as well as in the long wavelength limit) without a Rashba spin-orbit coupling is investigated. The remainder of the section is devoted to the calculation of the surface plasmon resonance (SPR) patches in the wave vector-frequency space is obtained solving the equation Real $\left(\varepsilon(\omega,|\delta \mathbf{k}|)\right) = 0$ . While our results and discussions are presented in the Sec.IV, there is a brief conclusion in Sec.V.

**2. Casimir-Polder Interaction** Ever since Hendrik Casimir found that two mirrors placed facing each other in a vacuum would attract, since then this bizarre force has been the object of intense experimental and theoretical scrutiny. In this section we shall first present a review of the calculation of the Casimir-Polder free energy corresponding to interactions of an electrically and magnetically polarizable micro-particle with a magneto-dielectric sheet in terms of the Fresnel coefficients of vacuum and the sheet material. This will be followed by the description of the gapped silicene/germanene system.

**A. Interactions of a micro-particle with a magneto-dielectric sheet** The Casimir (CI)and Casimir-Polder (CPI)interactions(whereas CI refers to force between two bulk objects, such as dielectric plates, CPI describes the force between a bulk object and a gas-phase atom)are conservative and arise due to the quantum fluctuations of the photon(electromagnetic) field or, more generally, from the zero-point energy of materials [33] and their dependence on the boundary conditions of the photon fields. The Casimir effect due to the vacuum fluctuations of the phonon field, similar to that due to the vacuum fluctuations of photon field, was also predicted for a Bose-Einstein condensate (BEC)[35]. A unified picture of these quantum-mechanical, fluctuation-driven-forces was given by Lifshitz [16,17,18] many decades ago: We consider a micro-particle in vacuum where the former is characterized by the dynamic electric polarizability $\xi_{microparticle}(\omega)$ and the dynamic magnetic polarizability $\eta_{microparticle}(\omega)$ as shown in Figure 1. The sample in the figure consists of a thin silicene film of thickness '*D*' deposited on a thick substrate at temperature *T*. Suppose the particle is at a separation '*d*' ($d >> d(T) \equiv \hbar c/(2k_B T)$ or, $T >> T_c \equiv \hbar c/(2dk_B)$) above the sample. Our aim is to investigate the interaction of this micro-particle with the given film material. To elucidate ab initio the concept of the classical Casimir-Polder interaction (CPI), leaning on the basis of the Lifshitz theory [16,17,18], we assume the interaction of the particle with the sheet to be of the *classical* CPI type in the first approximation. The classical case is valid when the separation is not small. The quantities $d(T)$ and $T_c$ set the classical limit in the sense that the limit starts from $d \approx 5\, d(T)$ and $T \approx 5T_c$. To explain, for ordinary materials at room temperature (300 K), $d(T) = \hbar c/(2k_B T)) \approx 3.66$ µm. Thus, the classical limit is achieved in this case approximately at separations above the edge $d = 10$ µm. In terms of temperature, for the separation $d = 10$ µm, the classical limit edge is $T_c \approx 110$ K. Thus, at $T >> 110$ K, say, at room temperature the classical limit is nearly achieved.

The projection of the wave vector on the (*x*,*y*) plane is denoted by $k^\perp$, and $k^\perp = |k^\perp|$. As we have already stated, we choose the coordinate plane (*x*, *y*) coinciding with the upper film surface and the z-axis perpendicular to it (Figure 1). The quantities $\varepsilon^{(0)}(i\omega_l) \to \varepsilon^{(0)}_{intervening\ medim}(i\omega_l)$ and $\mu^{(0)}(i\omega_l) \to \mu^{(0)}_{intervening\ medium}(i\omega_l)$ are the dynamic dielectric (relative) permittivity and the dynamic magnetic (relative) permeability of the intervening medium. If the medium happens to be vacuum and then static counterpart of each of them is equal to one. The micro-particle is characterized by the static polarizabilities, such as the electric polarizability $\xi_{microparticle}(0)$ and the magnetic polarizability $\eta_{microparticle}(0)$. The (electric) dipole polarizabilities are given in a variety of units, depending on the context in which they are determined. The most widely used unit for theoretical atomic physics is atomic units (a.u.), in which, e, $m_e$, the reduced Planck constant ℏ, and $4\pi\varepsilon_0$ have the numerical value 1. The quantity $4\pi\varepsilon_0 = 1$ implies that the electric polarizability in *a.u.* has the dimension of volume. However, the SI unit is C-m$^2$/V. The magnetic polarizability tensor($\beta_{ij}$) relates the induced magnetic dipole moment $m_i$ to the inducing field $B_j$ according to the equation $m_i = \beta_{ij} B_j$. The magnetic polarizability is also defined by the spin interactions of nucleons. Both the definitions lead to the SI unit $C^2$-m$^2$- kg$^{-1}$. We introduce now the polarizabilities $(\eta(0),\xi(0))$ in the atomic units. It may be mentioned that the polarizability ratio $Z_{2,\ microparticle} = (\eta_{microparticle} / \xi_{microparticle})^{1/2}$, though, has the SI unit m-s$^{-1}$, its dimensionless counterpart is $Z_2 = (\eta(0) / \xi(0))^{1/2}$ in a.u.. We shall use the polarizabilities in a.u. below in defining the free energy in Eq.(5). This choice has the great advantage of not have to worry about the dimensions of the various quantities involved in the numerics and the graphics to follow. Suppose the film material is characterized by the dielectric (relative) permittivity $\varepsilon^{(1)}(\omega)$ and the magnetic (relative) permeability $\mu^{(1)}(\omega)$, and the substrate is by the relative permittivity $\varepsilon^{(2)}(\omega)$ and the relative permeability $\mu^{(2)}(\omega)$. These might be made of either dielectric or metallic materials or poor conductor. We further assume that for the film material there exist finite limiting values of the relative permittivity and the permeability: $\varepsilon^{(1)}(0) \equiv \varepsilon_0^{(1)}$ and $\mu^{(1)}(0) \equiv \mu_0^{(1)}$. Correspondingly, the dimensionless impedance of the film material is $Z_1 = (\mu_0^{(1)}/\varepsilon_0^{(1)})^{1/2}$. If the film happens to be on a substrate then the static counterparts of $\varepsilon^{(2)}(i\omega_l)$ and $\mu^{(2)}(i\omega_l)$ are not equal to 1; the quantities $\varepsilon^{(2)}(i\omega_l)$ and $\mu^{(2)}(i\omega_l)$ are the dynamic dielectric permittivity and the magnetic permeability of the substrate material, respectively. Suppose, the finite limiting values of the static counterparts these quantities are $\varepsilon_0^{(2)}$ and $\mu_0^{(2)}$. If the film happens to be isolated, then $\varepsilon_0^{(2)}$ and $\mu_0^{(2)} = 1$.

We denote the reflection coefficients of the electromagnetic fluctuations on the sheet material plus substrate, dependent on the wave vector projection on the $(x,y)$ plane only and independent of frequency, for two independent modes, viz. the transverse magnetic (TM) and the transverse electric (TE) polarizations, by $\Pi_{tm}(0,k^\perp)$, and $\Pi_{te}(0,k^\perp)$, respectively. The closed-form, precise expressions for these reflection coefficients are given by the Fresnel coefficients $n_{tm}^{(n,n')}(i\omega_l, k^\perp)$, and $n_{te}^{(n,n')}(i\omega_l, k^\perp)$ **[9,10,14,15]** corresponding to the reflection on the boundary planes between the vacuum and the film material $(n = 0, n' = 1)$ and also between the film material and the substrate $(n=1, n'=2)$ :

$$\Pi_\alpha(0,k^\perp) = (n_\alpha^{(0,1)}(0,k^\perp) + n_\alpha^{(1,2)}(0, k^\perp) \exp(-2k^{(1)\perp}D)) / N_\alpha(0,k^\perp) \quad (1)$$

$$N_\alpha(0,k^\perp) = (1 + n_\alpha^{(0,1)}(0,k^\perp) n_\alpha^{(1,2)}(0,k^\perp) \exp(-2k^{(1)\perp}D)), \; \alpha = (t_m, t_e). \quad (2)$$

The Fresnel coefficients are calculated along the imaginary axis[9,10,14,15] and are given by

$$n_\alpha^{(n,n')}(i\omega_l, k^\perp) = [a^{(n')}(i\omega_l)k^{(n)}(i\omega_l, k^\perp) - a^{(n)}(i\omega_l)k^{(n')}(i\omega_l, k^\perp)]$$
$$/ [a^{(n')}(i\omega_l)k^{(n)}(i\omega_l, k^\perp) + a^{(n)}(i\omega_l)k^{(n')}(i\omega_l, k^\perp)], \quad (3)$$

where $a = (\varepsilon, \mu)$, $\alpha = (t_m, t_e)$, and $i\omega_l = (2\pi i l k_B T/\hbar)$ are (imaginary) Matsubara frequencies and

$$k^{(n)}(i\omega_l, k^\perp) = [k^{\perp 2} + \varepsilon^{(n)}(i\omega_l) \mu^{(n)}(i\omega_l)(\omega_l^2/c^2)]^{1/2}. \quad (4)$$

The dielectric constants $\varepsilon^{(n)}(i\omega_l)$, $\eta_\alpha(i\omega_l)$, etc. though written as function of frequency above, in general, is a function of frequency and the wave vector both. It describes the response of a medium to any field. For the fields slowly varying in space and time the limiting value of this function is the Faraday-Maxwell dielectric constant. So, under the assumption that the fields are slowly varying and temperature is near zero, the $l = 0$ term will only survive in Eq. (1). The configuration in Figure 1 comprises of an electrically and magnetically polarizable micro-particle, and a magneto-dielectric thin film. For this configuration in the classical limit **[9,10,14,15,16,17,18]**, upon considering only the $l = 0$ term and introducing a new (dimensionless) variable in place of $k^\perp$, viz. $\kappa \equiv 2d\, k^\perp$, the simple expressions for the Casimir-Polder free energy and the corresponding force, respectively, are given by

$$F(d,T) = - (k_B T f(d)/8d^3), \; \acute{K}(d,T) = - (3k_B T f(d)/8d^4) + (k_B T f'(d)/8d^3), \quad (5)$$

where

$$f(d) \equiv [_0\!\int^\infty \kappa^2 \, d\kappa \, e^{-\kappa} \times \{\xi(0) \Pi_{tm}(0, \kappa) + \eta(0) \Pi_{te}(0, \kappa)\}], \quad (6)$$

$$\Pi_{tm}(0,\kappa) = n_{tm}^{(0,1)}(0, \kappa) ( 1 - \exp(-\kappa D/d))/(1 - n_{tm}^{(0,1)}(0, \kappa)^2 \exp(-\kappa D/d)), \quad (7)$$

$$\Pi_{te}(0,\kappa) = n_{te}^{(0,1)}(0, \kappa) ( 1 - \exp(-\kappa D/d))/(1 - n_{te}^{(0,1)}(0, \kappa)^2 \exp(-\kappa D/d)). \quad (8)$$

$$n_{tm}^{(0,1)}(0, \kappa) = - n_{tm}^{(1,2)}(0, \kappa) = [\varepsilon_0^{(1)} - 1]/[\varepsilon_0^{(1)} + 1], \quad (9)$$

$$n_{te}^{(0,1)}(0, \kappa) = - n_{te}^{(1,2)}(0, \kappa) = [\mu_0^{(1)} - 1]/[\mu_0^{(1)} + 1]. \quad (10)$$

Since the quantities $\kappa \equiv 2d\, k^\perp$, and $(\Pi_{tm}(0,\kappa), \Pi_{te}(0, \kappa))$, as clear from (6) to (10), are dimensionless, the free energy function $(f(d)/d^3)$ is also dimensionless in a.u.. One has to now substitute the values of the dimensionless constants, such as $\varepsilon_0^{(1)}$ and $\mu_0^{(1)}$, to calculate the reflection coefficients ($\Pi_{tm}(0,\kappa)$, $\Pi_{te}(0,\kappa)$) which will eventually yield $F(d,T)$ and $\acute{K}(d,T)$ in the general situation.

As regards limiting values, the limit $D/d >> 1$ does not make sense. In the large- separation limit $D << d$, we find to the leading order in $D/d$ ($<< 1$)

$$F(d,T) \approx - (3k_BT\, Д(0)D/16d^4),\ \acute{K}(d,T) = - (3k_BT\, Д(0)D/4d^5),$$

$$Д(0) = \{\xi(0)(\varepsilon_0^{(1)2}-1)/\varepsilon_0^{(1)}\} + \{\eta(0) \times (\mu_0^{(1)2}-1)/\mu_0^{(1)}\}. \tag{11}$$

These are expressions for the Casimir-Polder free energy and the Casimir-Polder force in the large- separation limit. We find that the latter, in this limit, is attractive.

In the limit $D/d \sim 1$, the free energy function $f(d)$ is given by

$$f(d) = {}_0\!\int^N \kappa^2\, d\kappa\, e^{-\kappa\xi(0)}\left(1 - \exp\left(-\kappa\frac{D}{d}\right)\right)\left[\frac{Z_2^2\{(Z_1^2-1/\varepsilon_0^{(1)})/(Z_1^2+1/\varepsilon_0^{(1)})\}}{\left(1-\left(\frac{Z_1^2-1/\varepsilon_0^{(1)}}{Z_1^2+1/\varepsilon_0^{(1)}}\right)^2 \exp\left(-\kappa\frac{D}{d}\right)\right)}\right.$$

$$\left. + \frac{\{(\varepsilon_0^{(1)}-1)/(\varepsilon_0^{(1)}+1)\}}{\left(1-\left(\frac{\varepsilon_0^{(1)}-1}{\varepsilon_0^{(1)}+1}\right)^2 \exp\left(-\kappa\frac{D}{d}\right)\right)}\right]. \tag{12}$$

We notice that the important quantities in (12) are the dimensionless electric and magnetic polarizabilities of the micro-particle ($\xi(0), \eta(0)$) and the relative dielectric permittivity and the magnetic permeability of film material ($\varepsilon_0^{(1)}, \mu_0^{(1)}$). The Casimir–Polder (CP) force given above is $\acute{K}(d,T) = - (k_BT/8D^4)(D/d)^4(3\,f(d) - f'(d)\,d)$, where the integral $f(d)$ is given by Eq.(12) in terms of the dimensionless quantities ($Z_1$, $Z_2$). After a lengthy algebra involving the expansion of the exponential functions in $f(d)$ and $f'(d)$, which is tedious but not difficult, we find that the CP force is $\acute{K}(d,T,\varepsilon_0^{(1)},Z_1,Z_2) = -(k_BT/8d^4)\,g\,(d,\varepsilon_0^{(1)},Z_1,Z_2)$ where

$$g\,(d,\varepsilon_0^{(1)},Z_1,Z_2) = [3\,f(d,\varepsilon_0^{(1)},Z_1,Z_2) - d\,f'(d,\varepsilon_0^{(1)},Z_1,Z_2)],$$

$$= 120\xi(0)\left(\frac{D}{d}\right)\tilde{I}_2\left(\frac{D}{d}\right)\left(\mathfrak{I}_m a_e^2 + \mathfrak{I}_e a_m^2 Z_2^2\right) \times \left\{p - \left(\frac{\left(\frac{D}{d}\right)\tilde{I}_1\left(\frac{D}{d}\right)}{\tilde{I}_2\left(\frac{D}{d}\right)}\right)\right\},$$

$$p \equiv \left(\frac{(a_e + a_m Z_2)^2)}{5(\mathfrak{I}_m a_e^2 + \mathfrak{I}_e a_m^2 Z_2^2)}\right),\ a_e = \mathfrak{I}_m/(1-\mathfrak{I}_m^2),\ a_m = \mathfrak{I}_e/(1-\mathfrak{I}_e^2),$$

$$\mathfrak{I}_m^{(n=0, n'=1)} = \left(\frac{\varepsilon_0^{(1)}-1}{\varepsilon_0^{(1)}+1}\right),\ \mathfrak{I}_e^{(n=0, n'=1)} = \left(\frac{\varepsilon_0^{(1)}Z_1^2-1}{\varepsilon_0^{(1)}Z_1^2+1}\right). \tag{13}$$

Here $\tilde{I}_1\left(\frac{D}{d}\right)$ and $\tilde{I}_2\left(\frac{D}{d}\right)$ are the two slowly convergent series:

$$\tilde{I}_1\left(\frac{D}{d}\right) = \left[1 - 6\left(\frac{D}{d}\right) + \left(\frac{49}{2}\right)\left(\frac{D}{d}\right)^2 - \cdots\right],$$

$$\tilde{I}_2\left(\frac{D}{d}\right) = \left[1 - \left(\frac{5}{2}\right)\left(\frac{D}{d}\right) + 5\left(\frac{D}{d}\right)^2 - \left(\frac{35}{4}\right)\left(\frac{D}{d}\right)^3 + \cdots\right]. \tag{14}$$

One immediately obtains a criterion for the attractive Casimir-Polder interaction to turn repulsive. As long as we have

$$\left(\frac{\left(\frac{D}{d}\right)\tilde{I}_1\left(\frac{D}{d}\right)}{\tilde{I}_2\left(\frac{D}{d}\right)}\right) < p, \tag{15a}$$

the function $g(d,\varepsilon_0^{(1)},Z_1,Z_2)$ is greater than zero, and therefore the force is attractive. When

$$\left(\frac{\left(\frac{D}{d}\right)\tilde{I}_1\left(\frac{D}{d}\right)}{\tilde{I}_2\left(\frac{D}{d}\right)}\right) > p, \tag{15b}$$

the force turns repulsive as $g(d,\varepsilon_0^{(1)},Z_1,Z_2) < 0$. The term within the parenthesis in the right-hand-side of the first equation in (13) is $\left\{ p\tilde{I}_2\left(\frac{D}{d}\right) - \left(\left(\frac{D}{d}\right)\tilde{I}_1\left(\frac{D}{d}\right)\right)\right\}$. Upon expanding upto the fifth order in powers of $\left(\frac{D}{d}\right)$, we find that

$$F(d,\varepsilon_0^{(1)},Z_1,Z_2) = \left\{ p\tilde{I}_2\left(\frac{D}{d}\right) - \left(\left(\frac{D}{d}\right)\tilde{I}_1\left(\frac{D}{d}\right)\right)\right\} = -(21p + 260.4)\left(\frac{D}{d}\right)^5$$

$$+ (14p + 84)\left(\frac{D}{d}\right)^4 - (8.75p + 24.5)\left(\frac{D}{d}\right)^3 + (5p + 6)\left(\frac{D}{d}\right)^2 - (2.5p + 1)\left(\frac{D}{d}\right) + p. \tag{16}$$

Thus, the criterion (15) could now be expressed as if the dimensionless quintic $F(d,\varepsilon_0^{(1)},Z_1,Z_2)$ is greater (less) than zero, the Casimir-Polder interaction is attractive(repulsive). Our calculation above pertains to the Faraday-Maxwell (static) limit, where the frequency dependence of all functions are ignored completely, resulting in the appearance of these conditions. These conditions are obeyed by non-natural materials as shown in section5. We postpone further discussion on this issue till then. As regards the silicene film, in Figure 2, we have shown a plot of the dimensionless Casimir-Polder Force, at a given temperature and the sheet thickness, as a function of the ratio of the film thickness and the separation between the micro-particle and the film. We find the expected result that the magnitude of the attractive force (and the free-energy) increases as the distance between the micro-particle and the sheet decreases. We also find that as the effective dielectric constant increases there would be decrease in the magnitude of the attractive force between the micro-particle and the silicene-film.

A fundamental question here is that does the force need to be attractive? The answer is yes. To explain, we first note the well-known **[10,11,12,13]** fact that the atoms and molecules acquire temporary dipole moments as the space between interacting bodies is inhabited by "virtual" particle–antiparticle pairs fated to get annihilated in the time $\sim (\Delta E)^{-1}$ where $\Delta E$ is the energy of the fluctuations. These particle–antiparticle pairs, acting as the dipoles, radiate electromagnetic fields outward, and the fields interact with and scatter from fluctuating dipoles in the same as well as other macroscopic bodies. The reason for the scattering from the fluctuating dipoles being that the radiated fields can propagate over considerably long spatial ranges. To explain further, we take the convenient example of two parallel conducting plates, packed with radiating dipoles, with radiations of longer wavelengths dominating generally. Now the ability of dipoles on one plate to radiate at long wavelengths to dipoles on the other plate compared to the directions away from the plates is severely jeopardized due to the dipole moment and conductivity related boundary conditions. As a result of which the radiations away from the plates will be of much larger intensity compared to those directed towards the other plate. In other words, the long-wavelength radiation has more opportunities to leave the pair of plates entirely than to go between the plates, Inevitably, the impulse opposite to the former would push each plate toward the other. Having explained why the force is generally attractive, we hope that the explanation and calculation above set the tenor of the discussions to follow. It will be relevant to add that, for conducting parallel flat plates separated by a distance *d*, this force per unit area has the magnitude $(\pi^2/240)(\frac{\hbar c}{d^4})$ as calculated in ref.[1]. The role of *c* above is to convert the electromagnetic mode wavelength to a frequency, while ℏ converts the frequency to an energy. The absence of electronic charge implies that the electromagnetic field does not couple to matter in the usual sense here. It is important to note that when the separation, *d*, is so small that the mode frequencies are higher than the plasma frequency (for a metal) or higher than the surface Plasmon resonances or SPR (for a dielectric) of the material used to make the plates, this result breaks down. In the next section, one of the aims is to look for SPR.

The Casimir-Polder free energy corresponding to interactions *Ḱ(d,T)* of an electrically and magnetically polarizable micro-particle with a magneto-dielectric sheet is written down above in terms of the Fresnel coefficients of vacuum and the sheet material. In fact, the expression for the Casimir-Polder interaction (CPI) given in Eqs. (13) and (15) could be used for calculating *CPI $_{Si/Ge}$* by putting the value of the 'effective' dielectric constant of the Si/Ge sample in place of $\varepsilon_0^{(1)}$ *(and* $\mu_0^{(1)}$*= 1)* in these equations. The static polarization *Љ(0,δk,ξ,s$_z$=±1)* ,or the dielectric function *ε(0,a|δk|)* leads to the required value of the effective dielectric constant. Obviously, to begin with, an analysis to single-loop order is required to be performed considering the influence of the many-body effect on the Coulomb interaction on our system with two "flavors" of mass-less Dirac fermions. We consider the instantaneous bare Coulomb interaction given by

$$U_0(\mathbf{r},t) = (e^2/ḱ)\ \delta(t)/r = [(e^2/ḱ)\int (\frac{d^2(a|\delta\mathbf{k}|)}{2\pi}\int(\frac{d\omega}{2\pi})\ e^{(i.\delta\mathbf{k}.\mathbf{r}-i\omega t)}\ (a|\delta\mathbf{k}|)^{-1}] \qquad (17)$$

where *ḱ* is the 'effective' background dielectric constant. The background dielectric constant is *ḱ = (ḱ$_a$ +ḱ$_b$) /2* — the average screening contribution of any dielectric above (*ḱ$_a$*) and below (*ḱ$_b$*) the sample. The inclusion of the many-body effect is carried out by the polarization *Љ(ω,a|δk|)* insertion **[1-9]**. One obtains the effective Coulomb interaction, with the inclusion of the many-body effect, as

$$U(\mathbf{r},t)=(e^2/ḱ)\int(\frac{d^2(a|\delta\mathbf{k}|)}{2\pi}\int(\frac{d\omega}{2\pi})\ e^{(i.\delta\mathbf{k}.\mathbf{r}-i\omega t)}\ D(\omega,a|\delta\mathbf{k}|) \qquad (18)$$

where the Coulomb propagator $D(\omega,a|\delta\mathbf{k}|) = (a|\delta\mathbf{k}|+(2\pi e^2/ḱ)Љ(\omega,a|\delta\mathbf{k}|))^{-1}$. The asymptotics of the static polarization $(2\pi e^2/ḱ)Љ(0,a|\delta\mathbf{k}|)=Q(0,a|\delta\mathbf{k}|)$ at large or small wave vectors, respectively, thus, determines entirely the asymptotical behavior of the screened, static potential at small or large distances. In the random phase approximation(RPA), the dynamical dielectric function is thus $\varepsilon(\omega,a|\delta\mathbf{k}|) = ḱ\ [1+ (a\ |\delta\mathbf{k}|)^{-1}Q(\omega,\ a\ |\delta\mathbf{k}|)]$. We note that at the core of our novel scheme we have the dielectric effect; the proverbial "virtual" particle–antiparticle pairs **[1,2,3,4,5]**, acting as the dipoles, radiate electromagnetic fields outward as well as position themselves in such a way that they partially weaken the field. Not only CPI, the dynamical polarization function $Q(\omega,a\ |\delta\mathbf{k}|)$, or the dielectric function $\varepsilon(\omega,a|\delta\mathbf{k}|)$, leads to many valuable information, such as optical conductivity, plasmon dispersion, and so on. As already stated, these functions have been investigated by previous workers covering systems, such as graphene with and without an energy gap **[1-9]**. In Sec. III,we consider the

full formal expression of polarization function for 2+1 dimensional fermions on the normal/ferro-magnetic silicene plane with a general form of the propagator of quasi-particles. The propagator includes the spin-split single- particle excitation spectrum, the gap function near the valleys **K** and **K′**, and so on. We find that the many-body effects considerably change the bare Coulomb potential by way of the dependence of the static polarization $Q(0,a|\delta \mathbf{k}|)$, or the dielectric function $\varepsilon(0,a|\delta \mathbf{k}|)$ on the real-spin, iso-spin, and the ratio of the potential due to a field applied perpendicular to the silicene plane and the intrinsic spin-orbit coupling. Upon replacing the finite limiting value $\varepsilon_0^{(1)}$ of the dielectric constant of the sample by the effective dielectric function $\varepsilon(0,a|\delta \mathbf{q}| \to 0)$ ( or better still $\varepsilon(0,0)$) in the expression for Ќ(d,T) in Eqs. (13) and (15) below, one finally obtains the effective Casimir-Polder interaction(CPI) of a micro-particle with the sample. Note that our scheme to calculate CPI is though approximate in the sense that a macroscopic constant $\varepsilon_0^{(1)}$ in the classical Casimir-Polder force expression is proposed to be replaced by a microscopic quantity $\varepsilon(0,a|\delta \mathbf{q}| \to 0)$ calculated by an analysis to single-loop order, the coupling constant $\alpha \equiv (e^2/\kappa)/(\hbar v_F) \ll 1$, never-the-less, indicates that analysis to two-loop order and beyond is possibly not required.

In an effort to further exemplify our scheme, we now write the zero magnetic field, short wave-length result [9,10], viz. $Q(0,a|\delta \mathbf{k}|\gg 1) \approx (2\pi e^2/\kappa)(2|\delta \mathbf{k}|/(4\hbar v_F))$. This short wave-length (large momentum) result has incompatibility with the overall scheme-requirement of ours as the model Hamiltonian we proceed with is a low-energy one. We never-the-less move forward with the result above for the demonstration sake. The static, screened coulomb interaction is given by $U(\mathbf{r},0) = [e^2 \int_0^\infty d(a|\delta \mathbf{k}|) J_0(|\delta \mathbf{k}|,r)/\varepsilon(0,0)]$, where $\varepsilon(0,0) = \kappa + (\pi e^2)/(\hbar v_F)$ and the zeroth order Bessel function of the first kind is $J_0(|\delta \mathbf{k}|,r) = \int_0^{2\pi} d\theta\, e^{i\delta kr \cos(\theta)}$. As already mentioned, in atomic units (a.u.), e, $m_e$, $\hbar$, and $4\pi\varepsilon_0$ have the numerical value 1. The free-standing silicene has lattice parameter of $a = 3.89$ Å. We use the following other parameters in SI unit: $\kappa = 4\pi\varepsilon_0\varepsilon_r = 12\times 1.1111\times 10^{-10}$ N$^{-1}$-m$^{-2}$-Coul$^2$ (The value $\varepsilon_r = 12$ corresponds to the case where a silicon monolayer is deposited by molecular beam epitaxy (MBE) on a LAO(111) surface[38].) and $v_F = 10^6$ m/s. The quantity $(\pi e^2)/(\hbar v_F) = (0.7963)\times 10^{-9}$ N$^{-1}$-m$^{-2}$-Coul$^2$. Therefore, at large momentum, the 'effective' dielectric constant of the sample is $(\varepsilon(0,0)/4\pi\varepsilon_0) = 19.1667$. The background effect is of the same order as the bare relative permittivity in this case. The outcome (as shown in Figure 2) is along expected lines: the magnitude of the Casimir-Polder force and the free-energy decreases monotonically as the distance between the micro-particle and the sheet increases and vice versa. We also find that as the effective dielectric constant increases there would be decrease in the magnitude of the attractive force. It may be mentioned that the outcome is also true for a similar honeycomb structured material Germanene.

**B. Low-energy spectrum of silicene** The LYFE (Liu-Yao-Feng-Ezawa) model [19,20,21,39] – a variant of the Kane-Mele Hamiltonian[34]- of the system, including both intrinsic($t'_{so}$) and Rashba spin-orbit couplings($t'_1, t'_2$), can be written down in momentum space. Suppose $a_{\delta \mathbf{k},s_z}$ and $b_{\delta \mathbf{k},s_z}$ (with $s_z = \uparrow\downarrow$) denote the fermion annihilation operators corresponding to the A and B sub-lattices. The dimensionless low-energy Hamitonian around a Dirac point ( the iso-spin index $\xi = \pm 1$) in the basis $(a_{\delta \mathbf{k}\uparrow}, b_{\delta \mathbf{k}\uparrow}, a_{\delta \mathbf{k}\downarrow}, b_{\delta \mathbf{k}\downarrow})$ in momentum space is then given by

$$H = \sum_{\delta \mathbf{k}} (a^\dagger_{\delta \mathbf{k}\uparrow}\ b^\dagger_{\delta \mathbf{k}\uparrow}\ a^\dagger_{\delta \mathbf{k}\downarrow}\ b^\dagger_{\delta \mathbf{k}\downarrow}) \hbar(\delta \mathbf{k}) \begin{pmatrix} a_{\delta \mathbf{k}\uparrow} \\ b_{\delta \mathbf{k}\uparrow} \\ a_{\delta \mathbf{k}\downarrow} \\ b_{\delta \mathbf{k}\downarrow} \end{pmatrix}, \qquad (19)$$

where

$$\frac{\hbar(\delta k)}{\left(\frac{\hbar v_F}{a_0}\right)} \approx [\xi\, ak_x\tau_x + ak_y\tau_y + \xi\tau_z h_{11} + \Delta_z\tau_z + M\sigma_z + T'_{RI}(\xi\delta k)], \tag{20}$$

$$h_{11} = [\,t'_{so}\sigma_z + t'_2(ak_y\sigma_x - ak_x\sigma_y)], \tag{21}$$

$$T'_{RI} = t'_1(\xi\tau_x\sigma_y - \tau_y\sigma_x)/2,\ \Delta_{soc} = t'_{so} = \frac{t_{so}}{\left(\frac{\hbar v_F}{a_0}\right)},\ \Delta_z = \ell_0 E'_z = \frac{\ell E_z}{\left(\frac{\hbar v_F}{a_0}\right)}, \tag{22}$$

$$t'_1 = t_1(E_z)/(\hbar v_F/a_0),\ t'_2 = t_2/\left(\frac{\hbar v_F}{a_0}\right),\ M = M'/\left(\frac{\hbar v_F}{a_0}\right), \tag{23}$$

the (unstrained) half of the height difference between A and B atoms in a unit cell ($\ell_0$) ~ 0.3Å, and the unstrained lattice constant ($a_0$) = 3.86Å. The parameter $t_1(E_z)$ satisfies $t_1(E_z=0) = 0$ and becomes of the order of 10 μeV at the critical electric field $E_c$. Here $t_2$ ~ 1 meV is the second Rashba spin-orbit coupling (RSOC). In the presence of $t_2$ the energy bands will involve real spin-mixing. In fact, each energy band is characterized by a spin polarization index $s_z$ (where $s_z = +1(-1)$ corresponds to spin-up(spin-down) states), a valley polarization index $\xi$, and $\lambda$ (where $\lambda = +1(-1)$ denotes the electron (hole) states). The matrices $\sigma_i$ and $\tau_i$, respectively, denote the Pauli matrices associated with the real spin and pseudo-spin of the Dirac electronic states. The exchange field '$M'$' may arise due to coupling to a ferromagnet (FM) such as depositing Fe atoms to the silicene surface or depositing silicene to an FM insulating substrate. Since the Clifford algebra has an intimate connection with a Dirac system, the low-energy effective Hamiltonian(1) matrix above may also be written in terms of the Dirac matrices obeying this algebra. As $(t'_1, t'_2) \ll 1$, we find

$$\hbar(\delta k) = \left(\frac{\hbar v_F}{a}\right) \times [\xi\,\overline{\gamma}^1 a\,\delta k_x + \overline{\gamma}^2 a\,\delta k_y] + \xi[\,\overline{\gamma}^3\Delta_z + \Delta_{soc}\,\overline{\gamma}^3\gamma^5] - M\gamma^5, \tag{24}$$

where $\gamma^0 = -i\,\gamma^0$, $\overline{\gamma}^1 = (\gamma^5\gamma^0\gamma^x)$, $\overline{\gamma}^2 = (\gamma^5\gamma^0\gamma^y)$, $\overline{\gamma}^3 = (\gamma^5\gamma^z\gamma^0)$, and $\overline{\gamma}^5 = i\,\overline{\gamma}^0\overline{\gamma}^1\overline{\gamma}^2\overline{\gamma}^3$. The matrices $\overline{\gamma}^j$ comply with the usual rules of Clifford algebra: $\overline{\gamma}^\mu\overline{\gamma}^\nu + \overline{\gamma}^\nu\overline{\gamma}^\mu = 2g^{\mu\nu}$ where μ,ν = 0,1,2,3 and $g^{\mu\nu}$ = diag{-1, 1, 1,1}. The 4×4 matrices $\gamma^\mu$ are given as $\gamma^0 = \begin{pmatrix} 0 & I_2 \\ I_2 & 0 \end{pmatrix}, \gamma^i = \begin{pmatrix} 0 & \sigma_i \\ -\sigma_i & 0 \end{pmatrix}, \gamma^5 = \begin{pmatrix} -I_2 & 0 \\ 0 & I_2 \end{pmatrix}$ in the Weyl representation. Including the effect of the non-magnetic impurities **[39]** and assuming that the two Dirac valleys **K** and **K'** are decoupled, it is possible to write down a phenomenological, minimal Hamiltonian matrix in the basis $\{a^\xi_{\delta k, s_z}, b^\xi_{\delta k, s_z}\}^T$ where only the pseudo-spin is in the foreground; the iso-spin(described by the index $\xi = \pm 1$) and the real spin (described by an index $s_z = \pm 1$) are in the background:

$$\hbar_{\text{reduced}}/\left(\frac{\hbar v_F}{a}\right) \approx \sum_{\delta k,\xi,s_z}(\,a^{\xi\dagger}_{\delta\mathbf{k},s_z}\ \ b^{\xi\dagger}_{\delta\mathbf{k},s_z})\ \left(\frac{H(\delta k)_{\xi,s_z}}{\left(\frac{\hbar v_F}{a}\right)}\right)\begin{pmatrix} a^\xi_{\delta\mathbf{k},s_z} \\ b^\xi_{\delta\mathbf{k},s_z}\end{pmatrix}, \tag{25a}$$

$$\frac{H(\delta k)_{\xi,s_z}}{\left(\frac{\hbar v_F}{a}\right)} = \begin{pmatrix}(\Delta_{\xi,s_z} - \mu'_{s_z}) & a\delta k_- \\ a\delta k_+ & (-\Delta_{\xi,s_z} - \mu'_{s_z})\end{pmatrix}. \tag{25b}$$

Here $\Delta_{\xi,s_z} = (\Delta_z + \xi s_z \Delta_{soc})$, and $\mu'_{s_z} = ((\mu/\left(\frac{\hbar v_F}{a_0}\right)) - s_z M)$ is the dimensionless chemical potential of the fermion number dependent on the spin index $s_z$. One notices that the dependence of the bands on the spin quantum number $s_z$ is twofold: on the one hand it changes the Fermi level $\mu'$ to $\mu'_{s_z}$ and on the other it influences the gap given by

$\Delta_{\xi,s_z}$ defined above. In the presence of the non-magnetic impurities, the eight bands $\mathcal{E}^{(M)}(a|\delta k|)$ for the ferromagnetic silicene are now given by

$$\mathcal{E}^{(M)}(a|\delta k|) = -\mu'_{s_z} + \lambda\left[\{(a|\delta k|)^2 + \Delta_{soc}^{(M)}(\xi, s_z, a|\delta k|, \acute{\Gamma}_k(V_0))^2\}\right]^{1/2}. \quad (26)$$

where $\acute{\Gamma}_k(V_0)$ is the quasi-particle life time(QPLT)[39] depending on the Fourier transform $V_0$ of the non-magnetic impurity potential and

$$\Delta_{soc}^{(M)}(\xi, s_z, a|\delta k|, \acute{\Gamma}_k(V_0)) \equiv [\Delta_{\xi,s_z}^2 - (1/16\,\acute{\Gamma}_k(V_0)^2)\{(a|\delta k|)^2/(\Delta_{\xi,s_z}^2 + (a|\delta k|)^2)\}]^{1/2}. \quad (27)$$

We see that the significant alteration in the expression for the spectral gap ($\Delta_{\xi,s_z}$) occurs due to the inclusion of the elastic scattering by the non-magnetic impurities. It may be noted that the combination of the indices $(\xi, s_z)$ as '$\xi s_z$' makes the gap function $\Delta_{\xi,s_z}$ appear as the sum or the difference of two terms. When $\xi$ and $s_z$ are of the same sign we have the sum whereas when they are of the opposite sign we have the difference. On the other hand, one notices that the dependence of the bands on the spin quantum number $s_z$ alone is such that it changes the Fermi level $\mu'$ to $\mu'_{s_z}$ through the quantity $s_z M$. These lead to a few special features when the evolution of the band dispersions takes place with the increase in the out-of plane applied electric field. The increase leads to the transition from the topological insulator(TI) phase to the band insulator (BI) phase with the spin-valley-polarized metal(SVPM) in between. The SVPM state is attained by the mass-less fermions. For example, we find that if the valley $K$ ($\xi = +1$) corresponds to the SVPM state with indices $s_z = -1$, $\lambda = \pm 1$, in the valley $K'(\xi = -1)$ the SVPM state is attained by the mass-less fermions of the bands with indices $s_z = +1$, $\lambda = \pm 1$. The valence and conduction bands touch at the charge neutrality point (where the Fermi level is located). The massive fermions are present in overwhelming majority deep inside the TI and BI states, are also present in the valley $K$ inhabiting the bands $s_z = +1$, $\lambda = \pm 1$ and in the valley $K'$ the bands $s_z = -1$, $\lambda = \pm 1$. However, at the valley-spin-locked metal state the (mass-less) electrons and holes, acting as the carriers, would be preponderant. They have indices as follow : at the $K$ valley, the spin index $s_z = -1$, and the particle-hole index $\lambda = \pm 1$; at the $K'$ valley, the spin index $s_z = +1$, and the particle $-$ hole index $\lambda = \pm 1$. All these remarkable results are summarized in the table I.

**3. Surface Plasmon Resonance** In this section, we examine the vacuum polarization due to the presence of the electromagnetic field around planar charged fermions in silicene where the space between the fermions, according to the quantum field theory, are inhabited by "virtual" particle–antiparticle pairs fated to get annihilated in the time specified by the energy-time uncertainty relation. These pairs, acting as the dipoles, position themselves in such a way that they partially weaken the field. Evidently, this is a dielectric effect. This many-body effect changes the dielectric constant of the sample considerably even at the level of the one-loop contribution (see Figure 4) to the photon propagator for the finite or zero chemical potential of the fermions in the silicene. We expect that one-loop polarization function induces physical effects, which are likely to be revealed in a monolayer silicene sample in the absence of a magnetic field. The expression for the dynamic polarization function $Q(i\omega_m, \delta q)$ at finite temperature and finite chemical potential is given in the Matsubara representation in refs.**[6,7,8,9]**. The expression involves a bubble of two Green's functions (single-loop approximation) as shown in Figure 4. In the non-relativistic approximation, we obtain

$$Q(i\omega_m, a\delta q) = \sum_{\xi = \pm 1}(2\pi e^2/\kappa)\beta^{-1}\sum_{n=-\infty}^{n=+\infty}\int \left(\frac{d^2(a\delta k)}{4\pi^2}\right) \times tr\{G^{(0)}{}_\xi(\delta k, i\nu_n)$$

$$G^{(0)}{}_\xi(\delta\mathbf{k}+\delta\mathbf{q}, i\nu_n + i\omega_m)\}, \qquad (28)$$

where 'a' is the lattice constant and $G^{(0)}{}_\xi(\delta\mathbf{k}, i\nu_n)$ denotes the Green's function matrix of the non-interacting particle near the $K_\xi$ valley. The function $g^{(0)}{}_{\xi,\mu\nu}(\delta\mathbf{k}, i\nu_n)$ denotes an element of $G^{(0)}{}_\xi(\delta\mathbf{k}, i\nu_n)$ with the spin indices of the operators involved in the element as $(\mu, \nu)$. The Matsubara frequencies are given by $z_m = i\omega_m \equiv 2mi\pi/\beta\hbar$ and $z_n = i\nu_n \equiv (2n+1)i\pi/\beta\hbar$. The trace is over the spin indices, i.e. over the index $s_z = \pm 1$ in the present problem. Since the RSOC terms in the original Hamiltonian, which give rise to the spin-flip, have been ignored, there is no off-diagonal element of the form $g^{(0)}{}_{\xi,s_z,s'_z}(\delta\mathbf{k}, i\omega_n)$ with $s_z \neq s'_z$ in the Green's function matrix $G^{(0)}{}_\xi(\delta\mathbf{k}, i\nu_n)$. Consequently, the trace over $s_z$ simply reduces to a sum over $s_z$ and the summand is the product of the Green's functions matrices

$$\begin{pmatrix} g^{(0)}{}_{\xi,s_z=+1,s_z=+1,}(\delta\mathbf{k}, i\nu_n) & 0 \\ 0 & g^{(0)}{}_{\xi,s_z=-1,s_z=-1,}(\delta\mathbf{k}, i\nu_n) \end{pmatrix}$$

$$\times \begin{pmatrix} g^{(0)}{}_{\xi,s_z=+1,s_z=+1,}(\delta\mathbf{k}+\delta\mathbf{q}, i\nu_n + i\omega_m) & 0 \\ 0 & g^{(0)}{}_{\xi,s_z=-1,s_z=-1,}(\delta\mathbf{k}+\delta\mathbf{q}, i\nu_n + i\omega_m) \end{pmatrix}.$$

It follows that

$$tr\{G^{(0)}{}_\xi(\delta\mathbf{k}, i\nu_n) G^{(0)}{}_\xi(\delta\mathbf{k}+\delta\mathbf{q}, i\nu_n + i\omega_m)\}$$

$$= \sum_{s_z} g^{(0)}{}_{\xi,s_z,s_z}(\delta\mathbf{k}, i\nu_n) g^{(0)}{}_{\xi,s_z,s_z}(\delta\mathbf{k}+\delta\mathbf{q}, i\nu_n + i\omega_m) \qquad (29)$$

where the expression for the propagator $g^{(0)}{}_{\xi,s_z,s_z}(\delta\mathbf{k}, i\omega_n) \equiv u^2{}_{\xi,s_z,\delta\mathbf{k}}[i\hbar\omega_n - \acute{\varepsilon}^{(U)}{}_{\xi,s_z}(\mu', \delta\mathbf{k})]^{-1} + v^2{}_{\xi,s_z,\delta\mathbf{k}}[i\hbar\omega_n - \acute{\varepsilon}^{(L)}{}_{\xi,s_z}(\mu', \delta\mathbf{k})]^{-1}$ which is the diagonal element and the Bogoliubov coherence factors ($u^2{}_{\xi,s_z,\delta\mathbf{k}}$, $v^2{}_{\xi,s_z,\delta\mathbf{k}}$) are given by

$$u^2{}_{\xi,s_z,\delta\mathbf{k}} = (1/2)[1 - (\Delta_{\xi,s_z}/\acute{\varepsilon}_{\xi,s_z}(\delta\mathbf{k}))], v^2{}_{\xi,s_z,\delta\mathbf{k}} = (1/2)[1 + (\Delta_{\xi,s_z}/\acute{\varepsilon}_{\xi,s_z}(\delta\mathbf{k}))], \qquad (30)$$

Here $\Delta_{\xi,s_z} = (\xi s_z \Delta_{soc} + \Delta_z)$, and $\acute{\varepsilon}_{\xi,s_z}(\delta\mathbf{k}) = \sqrt{[(a|\delta\mathbf{k}|)^2 + \Delta^2_{\xi,s_z}]}$. The Green's function calculated above is obtained through the usual equation of motion method. The single-particle spectrum is given by $\acute{\varepsilon}^{(U)}{}_{\xi,s_z}(\mu', \delta\mathbf{k}) = -\mu'_{s_z} + [(a|\delta\mathbf{k}|)^2 + \Delta^2_{\xi,s_z}]^{1/2}$ and $\acute{\varepsilon}^{(L)}{}_{\xi,s_z}(\mu', \delta\mathbf{k}) = -\mu'_{s_z} - [(a|\delta\mathbf{k}|)^2 + \Delta^2_{\xi,s_z}]^{1/2}$ where $\mu'_{s_z} = ((\mu + s_z M)/(\frac{\hbar v_F}{a}))$. We first focus our attention in the simpler case of the total absence of the non-magnetic impurities, where $\acute{\varepsilon}^{(\infty)}{}_{\xi,s_z}(\delta\mathbf{k}) = [(a|\delta\mathbf{k}|)^2 + \Delta^2_{\xi,s_z}]^{1/2}$, $\acute{\varepsilon}^{(U)}{}_{\xi,s_z} = -\mu'_{s_z} + \lambda\acute{\varepsilon}^{(\infty)}{}_{\xi,s_z}$ with $\lambda = +1$, and $\acute{\varepsilon}^{(L)}{}_{\xi,s_z} = -\mu'_{s_z} + \lambda\acute{\varepsilon}^{(\infty)}{}_{\xi,s_z}$ with $\lambda = -1$. The superscript ($\infty$) has been introduced to emphasize that, due to the absence of the non-magnetic impurities, the quasi-particle life time is infinite. The coherence factors ($u^2{}_{\xi,s_z,\delta\mathbf{k}}$, $v^2{}_{\xi,s_z,\delta\mathbf{k}}$) are now given by $u^2{}_{\xi,s_z,\delta\mathbf{k}} = (1/2)[1 - (\Delta_{\xi,s_z}/\acute{\varepsilon}^{(\infty)}{}_{\xi,s_z}(\delta\mathbf{k}))]$ and $v^2{}_{\xi,s_z,\delta\mathbf{k}} = (1/2)[1 + (\Delta_{\xi,s_z}/\acute{\varepsilon}^{(\infty)}{}_{\xi,s_z}(\delta\mathbf{k}))]$. We introduce the symbols $i\nu_n = z_n$, and $i\nu_n + i\omega_m = z_n + z_m$ and write the summand of the expression

$$\sum_{n=-\infty}^{n=+\infty} \int \left(\frac{d^2(a\delta\mathbf{k})}{2\pi^2}\right) \times \sum_{\xi,s_z} g^{(0)}{}_{\xi,s_z,s_z}(\delta\mathbf{k}, z_n) g^{(0)}{}_{\xi,s_z,s_z}(\delta\mathbf{k}+\delta\mathbf{q}, z_n + z_m)$$

$$=\sum_{n=-\infty}^{n=+\infty} \int \left(\frac{d^2(a\delta k)}{2\pi^2}\right) \times \sum_{\xi,s_z} \left\{\frac{(\hbar z_m + \hbar z_n + \mu'_{s_z})(\hbar z_n + \mu'_{s_z}) + \Im_{\xi,s_z,\delta k,\delta q}((a|\delta k|)(a|\delta k+\delta q|) + \Delta^2_{\xi,s_z})}{\left[(\hbar z_m + \hbar z_n + \mu'_{s_z})^2 - \acute{\varepsilon}^{(\infty)2}_{\lambda,\xi,s_z}(\delta k+\delta q)\right]\left[(\hbar z_n + \mu'_{s_z})^2 - \acute{\varepsilon}^{(\infty)2}_{\lambda',\xi,s_z}(\delta k)\right]}\right\} (31)$$

where

$$\Im_{\xi,s_z,\delta k,\delta q} = (u^2_{\xi,s_z,\delta k} - v^2_{\xi,s_z,\delta k})(u^2_{\xi,s_z,\delta k+\delta q} - v^2_{\xi,s_z,\delta k+\delta q})$$

$$= \Delta^2_{\xi,s_z}/(\acute{\varepsilon}^{(\infty)}_{\xi,s_z}(\delta k)\,\acute{\varepsilon}^{(\infty)}_{\xi,s_z}(\delta k+\delta q)). \tag{32}$$

We now use the formula

$$\sum_{n=-\infty}^{n=+\infty} \{\hbar z_n + \mu'_{s_z}\mathcal{C}_1\}^{-1}\{\hbar z_m + \hbar z_n + \mu'_{s_z} + \mathcal{C}_2\}^{-1}$$

$$= \beta\{\check{n}_F(\mathcal{C}_1) - \check{n}_F(\mathcal{C}_2)\} \times \{\mathcal{C}_1 - \mathcal{C}_2 - \hbar z_m\}^{-1} \tag{33}$$

for the summation in Eq. (31). This is followed by the implementation of the analytic continuation from the Matsubara frequencies. The latter is made by the replacement $z_m \to (\omega + i0^+)$ required for retarded functions. We thus obtain

$$Q(\omega,\delta q) = \sum_{\xi,\lambda,\lambda'=\pm 1, s_z, s'_z} F_{\xi,\lambda,\lambda',s_z,s'_z} \left(\frac{2\pi e^2}{\acute{\kappa}}\right) \int \left(\frac{d^2(\hbar\delta k)D_\xi^{-1}}{4\pi^2}\right) \times$$

$$[\check{n}_F(\lambda\acute{\varepsilon}^{(\infty)}_{\xi,s_z}(\delta k)) - \check{n}_F(\lambda'\acute{\varepsilon}^{(\infty)}_{\xi,s_z}(\delta k + \delta q))] \tag{34}$$

for the retarded polarization function where the Coulomb propagator $D_\xi = \{\hbar\omega - \lambda\acute{\varepsilon}^{(\infty)}_{\xi,s_z}(\delta k) + \lambda'\acute{\varepsilon}^{(\infty)}_{\xi,s_z}(\delta k + \delta q) + i0^+\}$, the Fermi function $\check{n}_F(\varepsilon) = [exp(\beta(\varepsilon-\mu)) + 1]^{-1}$, and $\beta^{-1} = (k_BT)$. The overlap function

$$F_{\xi,\lambda,\lambda',s_z,s'_z} = (1/4)[1 + \lambda\lambda'\{\frac{((a|\delta k|)(a|\delta k+\delta q|) + \Delta^2_{\xi,s_z})}{\acute{\varepsilon}^{(\infty)}_{\lambda',\xi,s_z}(\delta k)\acute{\varepsilon}^{(\infty)}_{\lambda,\xi,s_z}(\delta k+\delta q)}\}] \tag{35}$$

comprises of the terms remaining in (31) after those required in writing $D^{-1}_\xi [\check{n}_F(\lambda\acute{\varepsilon}^{(\infty)}_{\xi,s_z}(\delta k)) - \check{n}_F(\lambda'\acute{\varepsilon}^{(\infty)}_{\xi,s_z}(\delta k + \delta q))]$ in the expression for $Q(\omega,\delta q)$ have been utilized. It may be seen that the Fermi functions $\check{n}_F(\varepsilon)$ turn into Heaviside step functions $\theta(\mu - \varepsilon)$ at zero temperature.

We have now the two cases when the effective chemical potential is less than the effective gap parameter ($\mu'_{s_z} < \Delta_{\xi,sz}$) and the effective chemical potential is greater than the effective gap parameter ($\mu'_{s_z} > \Delta_{\xi,sz}$). Consequently, in the zero-temperature limit, one may write the polarization function as $Q_{\xi,s_z}(\omega,\delta q,\mu'_{s_z},\Delta_{\xi,s_z}) = Q^{(0)}_{\xi,s_z}(\omega,\delta q,\mu'_{s_z},\Delta_{\xi,s_z})\,\Theta(\Delta_{\xi,sz} - \mu'_{s_z}) + Q^{(1)}_{\xi,s_z}(\omega,\delta q,\mu'_{s_z},\Delta_{\xi,s_z})\,\Theta(\mu'_{s_z} - \Delta_{\xi,sz})$. The latter case, where the Fermi level lies outside the gap, is basically corresponding to the long wave-length $((a|\delta q|) \ll 1)$ limit. In this limit the energy and momentum are small, and therefore the limit is compatible with the low energy description of the silicene given in subsection 2B. The real and the imaginary parts of the function $Q^{(1)}{}_{\xi,s_z}(\omega,\delta\boldsymbol{q},\mu'_{s_z},\Delta_{\xi,s_z})$ could be written down in the nine separate cases[8] of interest as in refs. [8]. Since these authors had gone through the exercise in detail, and the results are not more illuminating than what have been discussed for, we refrain from reproducing them again. We simply write the result for the single case, viz. $\hbar v_F|\delta k| \ll \hbar\omega \ll \mu$, we shall deal with. In this case

$$Q^{(1)}{}_{\xi,s_z}\left(\omega,\delta\boldsymbol{q},\mu'_{s_z},\Delta_{\xi,s_z}\right) = (2\mu'_{s_z}/\pi)\ \left[1-\left(\frac{\Delta_{\xi,s_z}}{\mu'_{s_z}}\right)^2\right]\left(\frac{(a|\delta q|)^2}{\left(\frac{a\omega}{v_F}\right)^2}\right)\left(\frac{\left(\frac{2\pi e^2}{\kappa a}\right)}{\left(\frac{\hbar v_F}{a}\right)}\right). \qquad (36)$$

Now, as has been mentioned in section 2, in the random phase approximation(RPA) the dynamical dielectric function $\varepsilon_{\xi,s_z}(\omega,a|\delta\boldsymbol{q}|,\mu'_{s_z},\Delta_{\xi,s_z})=[1+(a|\delta q|)^{-1}Q(\omega,a|\delta q|,\Delta_{\xi,s_z})]$, with background effect included. The long-wavelength limit dynamic dielectric function, for $\hbar v_F|\delta q| \ll \hbar\omega \ll \mu$, thus assumes the simple form

$$\varepsilon_{\xi,s_z}(\omega,a|\delta\boldsymbol{q}|,\mu'_{s_z},\Delta_{\xi,s_z}) = \acute{\kappa}\ \left[1-(2\mu'_{s_z}/\pi)\left\{1-\left(\frac{\Delta_{\xi,s_z}}{\mu'_{s_z}}\right)^2\right\}\frac{(a|\delta q|)}{\left(\frac{a\omega}{v_F}\right)^2}\times\left(\frac{\left(\frac{2\pi e^2}{\kappa a}\right)}{\left(\frac{\hbar v_F}{a}\right)}\right)\right]. \qquad (37)$$

The entire quantity $\left(\frac{\left(\frac{2\pi e^2}{\kappa a}\right)}{\left(\frac{\hbar v_F}{a}\right)}\right)\left(\frac{(a|\delta q|)}{\left(\frac{a\omega}{v_F}\right)^2}\right) < 1$, and the effective chemical potential of the fermion number $\mu'_{s_z} = ((\mu/\left(\frac{\hbar v_F}{a_0}\right)) - s_z M)$. The latter in the case of the ferromagnetic silicene has been redefined through the presence of the exchange field M. Furthermore, the topological phase transition(TPT) in silicene and the germanene is characterized by the gap function $\Delta_{\xi,s_z}$ approaching zero. It may be roughly concluded that the microscopic quantity $\varepsilon_{\xi,s_z}\left(\omega,a|\delta\boldsymbol{q}|,\mu'_{s_z},\Delta_{\xi,s_z}\right)$ for a normal silicene attains higher value while the silicene system approaches TPT due to the continuous tuning of the electric field. For a ferromagnetic silicene, theoretically, one encounters a spin-split scenario for the valleys **K** and **K′**, for, say, the valley **K** we have $\varepsilon_{\xi=+1,s_z=+1} < \varepsilon_{\xi=+1,s_z=-1}$. Since the quantity $\varepsilon_{\xi,s_z}\left(\omega,a|\delta\boldsymbol{q}|,\mu'_{s_z},\Delta_{\xi,s_z}\right)$ is not a suitable experimental option on account of its dependence on the various parameters, we look for the finger-print of TPT elsewhere, such as in the expression of its static counter-part in the situation when $a|\delta\boldsymbol{q}| \ll 1$. The appropriate function for this analysis is the dielectric function $\varepsilon_{\xi,s_z}\left(\omega=0,a|\delta\boldsymbol{q}|,\mu'_{s_z},\Delta_{\xi,\varepsilon,s_z}\right)$ given by

$$\varepsilon_{\xi,s_z}\left(\omega=0,a|\delta\boldsymbol{q}|,\mu'_{s_z},\Delta_{\xi,\varepsilon,s_z}\right) = \acute{\kappa}[1+\Theta(\mu'_{s_z}-\Delta_{\xi,\varepsilon,sz})\ (2\mu'_{s_z}/\pi)\ F^{(1)}{}_{\xi,s_z}(\omega=0,\delta\boldsymbol{q},\mu'_{s_z},\Delta_{\xi,\varepsilon,s_z})]$$

(38)

where

$$F^{(1)}{}_{\xi,s_z}(\omega=0,\delta\boldsymbol{q},\mu'_{s_z},\Delta_{\xi,s_z}) = \left(\frac{\left(\frac{2\pi e^2}{\kappa a}\right)}{\left(\frac{\hbar v_F}{a}\right)}\right)(a|\delta q|)^{-1}\ [\Theta(2ak^F_{\xi,s_z,\varepsilon}-a\delta q)-\Theta(a\delta q-2ak^F_{\xi,s_z,\varepsilon})$$

$$\times\left\{\left(\frac{\sqrt{\{(a|\delta q|)^2-4ak^{F2}_{\xi,s_z,\varepsilon}\}}}{2(a|\delta q|)}\right)-\left(\frac{(a|\delta q|)^2-4\Delta^2_{\xi,\varepsilon,s_z}}{4\mu'_{s_z}(a|\delta q|)}\right)\times arctan\left(\frac{\sqrt{\{(a|\delta q|)^2-4ak^{F2}_{\xi,s_z,\varepsilon}\}}}{2\mu'_{s_z}}\right)\right\}]. \qquad (39)$$

We now consider the primary inequality $a\delta q < 2ak^F_{\xi,s_z,\varepsilon}$ (where the quantity $2ak^F_{\xi,s_z,\varepsilon} \equiv 2\sqrt{(\mu'^2_{s_z} - \Delta^2_{\xi,\varepsilon,s_z})}$) for which

$$\varepsilon_{\xi,s_z}\left(\omega = 0, a|\delta q|, \mu'_{s_z}, \Delta_{\xi,\varepsilon,s_z}\right) = \acute{\kappa}[1 + \Theta(\mu'_{s_z} - \Delta_{\xi,\varepsilon,sz})\,(2\mu'_{s_z}/\pi)\left(\frac{\left(\frac{2\pi e^2}{\acute{\kappa} a}\right)}{\left(\frac{\hbar v_F}{a}\right)}\right)(a|\delta q|)^{-1}] \quad (40)$$

with $\acute{\kappa} = 4\pi\varepsilon_0\varepsilon_r = 12\times 1.1111\times 10^{-10}$ N$^{-1}$-m$^{-2}$-Coul$^2$. Since the gap function is $\Delta_{\xi,\varepsilon,s_z} \sim 8$ meV, the quantity μ has to be greater than 8 meV here. The dimensionless quantity $(\mu/\left(\frac{\hbar v_F}{a_0}\right)) = 0.0108$ and therefore $\mu'_{s_z=\pm 1} = 0.0108 \mp M$. For M = 0, thus the quantity $2ak^F_{\xi,s_z,\varepsilon} \equiv 2\sqrt{(\mu'^2_{s_z} - \Delta^2_{\xi,\varepsilon,s_z})} \approx 0.0183$. However, as TPT is approached, $2ak^F_{\xi,s_z,\varepsilon} \approx 0.0216$. For the non-zero M, say M = 0.005, the quantity $2ak^F_{\xi,s_z,\varepsilon}$ at TPT is $2ak^F_{\xi,s_z,\varepsilon} \approx 0.0116$ and 0.0316, respectively, for $s_z = +1$ and $s_z = -1$. Away from TPT, $2ak^F_{\xi,s_z,\varepsilon} \approx 0.0011$ and 0.0294, respectively, for $s_z = +1$ and $s_z = -1$. Assuming that one could be as close to the Dirac neutrality point as possible, the quantity $(a|\delta q|)$ could be ensured less than $2ak^F_{\xi,s_z,\varepsilon}$. We witness an interesting situation characterized by (i) it is possible to have different dielectric response corresponding to the down and the up spin over a wide range of the exchange field in the presence of the ferromagnetic impurities, and (ii) in our scheme of the CPI calculation the upper limit of the integration in (12) and (13) for the normal silicene is 0.0183 elsewhere but as the TPT is approached the upper limit is 0.0216. For the ferromagnetic silicene, these limits are (0.0011, 0.0294) elsewhere and (0.0116, 0.0316) at TPT. The ramification of these results on the CPI involving the ferromagnetic silicene will be qualitatively similar. We now plot the dimensionless Casimir-Polder Force (CPF) given by Eqs. (12) and (13), at a given temperature and the sheet thickness, as a function of the ratio of the film thickness and the separation between the micro-particle and the film in Figure 2(b). The curve 1 corresponds to the case when the system is at TPT and the curve 2 when the system is at the topological/trivial insulating phase. We find that the magnitude of the force at a given ratio of the film thickness and the separation between the micro-particle and the film is greater at TPT than at the topological insulator and trivial insulator phases. The 2D plots of the real and the imaginary parts of the static dielectric function, on the other hand, as a function of the dimensionless wave number $(a\delta q)$ in the limit $a\delta q > 2ak^F_{\xi,s_z,\varepsilon}$ have been shown in Figure 3. Since the real and the imaginary parts tend to zero as $a\delta q \to 0.2$ in Figure 3, it is clear from these curves that the upper limit of the momentum integration in Eq.(16) may be fixed at 0.2 to deal with the case $a\delta q > 2ak^F_{\xi,s_z,\varepsilon}$. We digress for a while presently and introduce the biaxial strain field $(\varepsilon)$ and its influence, particularly, in the band gap $\Delta_{\xi,s_z}$ and half of the height difference $\ell$ between A and B atoms in a unit cell. This is yet another field which is tunable and leaves a discernable impact on TPT**[40]**.

There is a growing realization that mechanical strain and curvature leave important finger-print **[41, 42, 43]** on the charge and the spin related properties of the fermions even on the buckled honey-comb structured plane. Accordingly, we try to map out the effect of strain field on these important properties, such as the surface plasmon resonance frequency. We note that, as the gap function is dependent on the electric field $E_z$, the dielectric function $\varepsilon_{\xi,\varepsilon,s_z}$ is electrically tunable. The plasmon branch can be obtained by equating real part of

$\varepsilon_{\xi,\varepsilon,s_z}$ with zero. It can also be obtained by finding the zeros of the dielectric function $\varepsilon_{\xi,\varepsilon,s_z}$ ($\omega_p - i\gamma$, $\delta\mathbf{q}$), where $\gamma$ is the plasmon decay rate. The imaginary part of the polarization function plays an important role in determining the behavior of the plasmons. The regions in ($\delta q$, $\omega$) space of non-zero function correspond to regions in which collective oscillations are damped. This is also referred to as the Landau damping. The long-wavelength limit polarization for the system under consideration has been exhaustively investigated recently by Iurov et al.[44] at the finite-temperature. We find that in the case of the strained system, for example, the plasmon frequency in the long wave-length limit (and for the massive fermions) will be given by

$$\left(\frac{a\omega_{massive}}{v_F}\right)^2 = 4(a|\delta\boldsymbol{q}|)[\log_e 2 \; \frac{k_BT}{\frac{\hbar v_F}{a}} - (\Delta^2_{\xi,s_z,\varepsilon} / \frac{4k_BT}{\frac{\hbar v_F}{a}}) \{1 + \log_e \left[\frac{\Delta_{\xi s_z\varepsilon}}{\frac{2k_BT}{\left(\frac{\hbar v_F}{a}\right)}}\right]\}] \times \left(\frac{\left(\frac{2\pi e^2}{\kappa a}\right)}{\left(\frac{\hbar v_F}{a}\right)}\right).$$

(41)

The strain is defined [15] as $\varepsilon = (a - a_0)/a_0$, where $a$ and $a_0$ are strained and unstrained lattice-constants respectively. The quantity $\Delta_{\xi,s_z,\varepsilon}$ above is equal to $2 |\ell(\varepsilon)E'_z + \xi\, s_z\, \Delta_{soc}(\varepsilon)|$ where $\Delta_{soc}(\varepsilon) = \Delta_{\varepsilon,E_z=0}/\left(\frac{\hbar v_F}{a_0}\right)$ and $\ell(\varepsilon)$-the (strained) half of the height difference between A and B atoms in a unit cell. The mass-less fermions have simpler expression for the long-wave-length limit plasmon frequency:

$$\left(\frac{a\omega_{massless}}{v_F}\right)^2 = 4(a|\delta\boldsymbol{q}|)[\log_e 2 \; \frac{k_BT}{\frac{\hbar v_F}{a}}] \times \frac{(2\pi e^2/\kappa a)}{\left(\frac{\hbar v_F}{a}\right)}.$$

(42)

At the finite-temperature, one thus finds the linear (T,$|\delta\boldsymbol{q}|$) dependence of the squared plasmon frequency. Iurov et al.[44] have found non-linear(T, $|\delta\boldsymbol{q}|$) dependence of the damping rates. As already mentioned, the emerging trend is to look for the signature of the mechanical strain on the charge and the spin related properties of the fermions as mentioned above. To set the stage to map out the effect of the strain field, we apply a hydrostatic biaxial strain($\varepsilon$) on the silicene layer. We shall see below that the critical field strength $|E_{zc}'|$ increases with compressive and tensile strain field. It is well-known [40] that the decrease in the tensile stain ($\varepsilon > 0$) and the increase in the absolute magnitude ($|\varepsilon|$) of the compressive strain($\varepsilon < 0$) brings about an increase in the band-gap $\Delta_{\varepsilon,E_z=0}$ of the silicene where the gap function is to be made dimensionless dividing by $\left(\frac{\hbar v_F}{a_0}\right) = 1.3856\, A^0$ as before. In the presence of the strain field, the LYFE model [4,5,6,13,14,15] result for the gap function, viz. $G_{\xi,s_z}(E_z, \varepsilon = 0) = 2|\Delta_{\xi,s_z}|$ where $\Delta_{\xi,s_z} = (\Delta_z + \xi\, s_z\, \Delta_{soc})$ and $\Delta_z = \ell_0 E'_z$, could be generalized as $G_{\xi,s_z}(E_z, \varepsilon) = 2\,|\ell(\varepsilon)E'_z + \xi\, s_z\, \Delta_{soc}(\varepsilon)| \sim 10$ meV where $\Delta_{soc}(\varepsilon) = \Delta_{\varepsilon,E_z=0}/\left(\frac{\hbar v_F}{a_0}\right)$. Upon using the set of data in ref.[15] relating the gap function $\Delta_{\varepsilon,E_z=0}$ as a function of the hydrostatic, biaxial strain field($\varepsilon$), we find by the method of least squares that the dimensionless gap function is given by $\Delta_{\varepsilon,E_z=0}/\left(\frac{\hbar v_F}{a_0}\right) = 0.0011 - 0.0153\varepsilon + 0.0879\varepsilon^2 + 0.4992\varepsilon^3 - 3.1053\varepsilon^4 + 7.0260\varepsilon^5$ as shown in Figure 5(a). Another version of this result is $\Delta_{\varepsilon,E_z=0}/\left(\frac{\hbar v_F}{a_0}\right) = -0.0231 + 0.1241p - 0.2527p^2 + 0.2490p^3 - 0.1117p^4 + 0.0215p^5$ where $p(\varepsilon) = \ell(\varepsilon)/\ell_0$, $\ell_0$ is the half of the height difference between A and B atoms in a unit cell in the absence of the strain field, and $\ell(\varepsilon)$-the (strained) half of the height difference between A and B atoms. This half height difference ratio depends on $\varepsilon$ as $p(\varepsilon)=(1-9\varepsilon-4\varepsilon^2+492\,\varepsilon^3 + 7584\,\varepsilon^4 - 3690.4\,\varepsilon^5)$. In the figure 5(b) above we have 3D-plotted

$G_{\xi,s_z}$ ($E_z$, $\varepsilon$) as a function of $E_z$ and $\varepsilon$ for $\xi = +1$ and $s_z = -1$. The figure shows that the critical field $E'_{zc} = -\xi s_z \frac{\Delta_{soc}(\varepsilon)}{\ell(\varepsilon)}$, at which the gap function $G_{\xi,s_z}$ ($E_z$, $\varepsilon$) becomes zero, is a function of the strain field : $E'_{zc} = E'_{zc}(\varepsilon)$. Thus, the strain field affects TPT to the extent of the shift in the value of the critical field. The 3D-plot of $G_{\xi,s_z}$ ($E_z$, $\varepsilon$) with the characteristic W-shape[16], in fact, corresponds to the phase diagram of the system in the $\varepsilon$ - $E_z$ plane. We, further, find that for a given value of $\varepsilon$, $G_{\xi,s_z}$ ($E_z$, $\varepsilon$) decreases to zero as |$E'_z$| increases from zero to |$E'_{zc}$|. The gap function increases now, attains peak value, and decreases to zero as the electric field varies as $-E'_{zc} \leq E'_z \leq E'_{zc}$. The three regions traversed in the process are band insulator (region I) region, valley-spin-locked metal, or VSLM, strip (region II), and the quantum spin Hall insulator (QSH)state. Thus the figure 5(b) clearly shows that the critical field strength |$E_{zc}'$| increases with compressive and tensile strain. As regards Figure 5(c), we have plotted here $\left(\frac{a\omega}{v_F}\right)$ as a function of $\varepsilon$ for $\xi = +1$ and $s_z = -1$ as given in Eq.(40) at the room temperature T = 300 K. Save the dependence that the gap function $G_{\xi,s_z}$ ($E_z$, $\varepsilon$) requires an infinitesimal electric field to become a real number, at a given wave vector $(a|\delta q|) \ll 1$, we notice that the squared plasma frequency $\left(\frac{a\omega_{massive}}{v_F}\right)^2$ or $\left(\frac{a\omega_{massless}}{v_F}\right)^2$ for the massless as well as the massive fermions does not get influenced by the variation in the electric field $E_z$ at non-zero temperature. The quantity $\left(\frac{a\omega}{v_F}\right)$, however, depends on the strain field linearly as could be seen in Figure 5(c). At the room temperature T = 300 K, the plots of $\left(\frac{a\omega}{v_F}\right)$ in this figure are for the massive as well as the mass-less fermions as given by Eqs.(41) and (42), respectively. The figure shows that the decrease in the tensile stain ($\varepsilon > 0$) and the increase in the absolute magnitude (|$\varepsilon$|) of the compressive strain($\varepsilon < 0$) brings about a decrease in the Plasmon frequency for the mass-less as well as the massive fermions in the long wavelength limit. We find that the valley-spin-locked (VSL) phase (represented by the upper curve in figure 5(c)) is characterized by the higher plasmon frequency for both the varieties of fermions compared to the topologically/trivially insulating phases represented by the lower curve in figure 5(c).

The case, $\mu'_{s_z} < \Delta_{\xi,s_z,\varepsilon}$, is relevant and worth-examining as it corresponds to the intrinsic limit. The zero-temperature polarization function in the strained case is given by

$$Q_{\xi,s_z}(\omega, \delta q, \mu'_{s_z}, \Delta_{\xi,s_z}) = (2\mu'_{s_z}/\pi) \left\{\frac{(a|\delta q|)^2}{(a|\delta q|)^2 - \left(\frac{a\omega}{v_F}\right)^2}\right\} \times \left[\frac{\pi\Delta_{\xi,s_z,\varepsilon}}{\mu'_{s_z}} + \left\{\frac{\left((a|\delta q|)^2 - \left(\frac{a\omega}{v_F}\right)^2 - 4\Delta^2_{\xi,s_z,\varepsilon}\right)}{\left(\frac{2\mu'_{s_z}}{\pi}\right)\sqrt{\{(a|\delta q|)^2 - \left(\frac{a\omega}{v_F}\right)^2\}}}\right\}\right.$$

$$\left. \times \arcsin\sqrt{\left(\{(a|\delta q|)^2 - \left(\frac{a\omega}{v_F}\right)^2\}/\{(a|\delta q|)^2 - \left(\frac{a\omega}{v_F}\right)^2 + 4\Delta^2_{\xi,s_z,\varepsilon}\}\right)}\right]. \quad (43)$$

In the static case, the polarization function is given by $Q_{\xi,s_z}(\omega = 0, \delta q, \mu'_{s_z}, \Delta_{\xi,s_z}) = (2\mu'_{s_z}/\pi) [F^{(0)}{}_{\xi,s_z}(\omega = 0, \delta q, \mu'_{s_z}, \Delta_{\xi,s_z}) \Theta(\Delta_{\xi,sz} - \mu'_{s_z})]$ where

$$F^{(0)}{}_{\xi,s_z}(\omega = 0, \delta q, \mu'_{s_z}, \Delta_{\xi,s_z}) = \frac{\Delta_{\xi,s_z}}{2\mu'_{s_z}} + \left(\frac{(a|\delta q|)^2 - 4\Delta^2_{\xi,s_z}}{4\mu'_{s_z}(a|\delta q|)}\right) \times \arcsin\sqrt{\left(\frac{(a|\delta q|)^2}{\{(a|\delta q|)^2 + 4\Delta^2_{\xi,s_z,\varepsilon}\}}\right)}. \quad (44)$$

One may obtain the finite-temperature counterpart following Iurov et al.[44]. We, however, wish to analyze the simpler expression (43) as far as possible. This expression leads to the effective coulomb interaction, with the inclusion of the many-body effect, as follows: The instantaneous bare coulomb interaction is given by $U_0(r,t) = (e^2/\acute{\kappa})\,\delta(t)/r = [\int(\frac{d^2(\delta q)}{4\pi^2}\int(\frac{d\omega}{2\pi})\,e^{(i.\delta q.r-i\omega t)}\,U_0(|\delta q|)]$ where $D_0(\omega,|\delta q|) = U_0(|\delta q|) = (2\pi e^2/\acute{\kappa}a)(a|\delta q|)^{-1}$ is the bare coulomb propagator and $\acute{\kappa}$ is the effective background dielectric constant. The background dielectric constant $\acute{\kappa} = (\acute{\kappa}_a + \acute{\kappa}_b)/2$ is the average screening contribution of any dielectric above ($\acute{\kappa}_a$) and below ($\acute{\kappa}_b$) the sample. The additional screening of an instantaneous bare coulomb interaction, due to many-body effects, is determined by the retarded polarization function $Q(\omega,\delta q)$ or, equivalently by the dieletric function

$$\varepsilon(\omega,|\delta q|) = [1 + (a|\delta q|)^{-1}\,Q(\omega,\delta q)] = (a|\delta q|)^{-1}\,[(a|\delta q|)+Q(\omega,\delta q)]. \qquad (45)$$

This allows us to write the effective coulomb interaction $U(r,t) = [\int(\frac{d^2(\delta q)}{4\pi^2}\int(\frac{d\omega}{2\pi})\,e^{(i.\delta q.r-i\omega t)}\,D(\omega,|\delta q|)]$ where the effective coulomb propagator is $D(\omega,|\delta q|) = (2\pi e^2/\acute{\kappa}a)(a|\delta q| + Q(\omega,\delta q))^{-1} = U_0(|\delta q|)\,\varepsilon^{-1}(\omega,|\delta q|)$. The zero-temperature plasmon resonance frequency $\omega_p(k)$ is obtained within the RPA approximation by finding zeros of the dielectric function $\varepsilon(\omega,|\delta q|,\Delta_{\xi,s_z},T=0)$ i.e. by solving the equation

$$\varepsilon(\omega,|\delta q|,\Delta_{\xi,s_z},T=0) = \left[1 + \left\{\frac{(2\Delta_{\xi,s_z,\varepsilon})\,(a|\delta q|)}{(a|\delta q|)^2-\left(\frac{a\omega}{v_F}\right)^2}\right\} \times \left(1+\left\{\frac{(a|\delta q|)^2-\left(\frac{a\omega}{v_F}\right)^2-4\Delta^2_{\xi,s_z,\varepsilon}}{\left(2\Delta_{\xi,s_z,\varepsilon}\right)\sqrt{\{(a|\delta q|)^2-\left(\frac{a\omega}{v_F}\right)^2\}}}\right\}\right.\right.$$

$$\left.\left. \times\ arcsin\sqrt{\left(\{(a|\delta q|)^2-\left(\frac{a\omega}{v_F}\right)^2\}/\{(a|\delta q|)^2-\left(\frac{a\omega}{v_F}\right)^2+4\Delta^2_{\xi,s_z,\varepsilon}\}\right)}\right)\right] = 0. \qquad (46)$$

in the case $\mu'_{s_z} < \Delta_{\xi,s_z,\varepsilon}$. Under the assumption that the damping of the collective resonance mode is weak, we find from Eq.(45) and the 2 D graphical representations in Figure 6 that this resonance mode is present for the gapped (or massive) fermions only. In the Dirac limit, the intrinsic plasmons are absent. The Plasmon frequency, by and large, is of the order 10 THz. The contour plots of $\varepsilon(\omega,|\delta k|)$ in the (k, ω) space in the intrinsic limit(($\mu' < \Delta_{\xi,s_z}$) have been shown in Figure 7. The surface plasmon resonance (SPR) patches in the (δk, ω) space is obtained solving the equation Real $(\varepsilon(\omega,|\delta k|)) = 0$. We have divided the (δk, ω) space in the two regions, viz. the upper region $(a|\delta q|) \leq \left(\frac{a\omega}{v_F}\right)$ and the lower region $(a|\delta q|) \geq \left(\frac{a\omega}{v_F}\right)$. In Figure 7(a), the contour plot of $\varepsilon(\omega,|\delta k|)$ for the gapless fermions show that the SPR is absent in the upper as well as the lower regions. In Figure 7(b), for the massive fermions, the yellow patches in the (k, ω) space correspond to the real part of the polarization function zeros and therefore they form the so-called SPR region there. The SPR region extend to parts of the upper as well as lower regions. The electric field ratio $E_z/E_c = 1$ for both the cases. The Figures 7(c) and 7(d), respectively, show the contour plot of $\text{Im}(\varepsilon(\omega,|\delta k|))$ as a function of (k,ω) for the gapless and the gapped fermions with $E_z/E_c = 1$. We notice that the single-particle-excitation (SPE) continuum $(\text{Im}((\varepsilon(\omega,|\delta k|)) \neq 0)$ is everywhere except in the burnt-red patch. The decay of plasmons into electron-hole(e-h) pairs, known as the Landau damping, occurs in the SPE continuum. The contour plot of Re $(\varepsilon(\omega,|\delta k|))$ as a function of (k,ω) for the massive fermions for $E_z/E_c = 0.5$ ( $E_z/E_c = 1.5$) have been shown in Figure (7e)[7(f)]. In both the cases there is distinct possibility of the occurrence of the SPR as real $(\varepsilon(\omega,|\delta k|)) = 0$ will turn out to be zero in some patches. So, the noteworthy feature of 2+1 dimensional spin-orbit coupled fermions, which is underscored by findings in Figures(6) and (7), is that, though the gapless fermions in the VSL metallic phase completely shuns SPR, this is not true for the gapped counterpart.

**4. Results and Discussion** We note that our scheme to calculate CPI is slightly intrepid in the sense that, on the one hand, we started with a classical Casimir-Polder interaction expression where the reflection coefficients of the electromagnetic fluctuations on the silicene sheet plus substrate are given by the usual Fresnel coefficients corresponding to the reflection on the boundary planes between the vacuum and the film material and also between the film material and the substrate, on the other hand the quantum field theoretic methods have been deployed to replace a macroscopic constant $\varepsilon_0^{(1)}$, involved in writing the Fresnel coefficients, by a microscopic quantity $\varepsilon(0, a|\delta\boldsymbol{q}|\to 0)$. This allows us to introduce the Casimir-Polder free energy between a micro-particle and the silicene film. Using this expression, we numerically integrate to obtain the Casimir force as we tune our original Hamiltonian through a topological phase transition. Perhaps the consistency demands that we use a microscopic quantum field theoretic method to calculate the Casimir energy at zero temperature in terms of, say, the current-current correlation functions of the two electron systems under consideration and virtual photons in the 3D vacuum between them. A (softer) counterview is that our analytical and numerical calculations are an amalgamation of the classical ideas concerning the reflection coefficients of the electromagnetic fluctuations on the silicene sheet plus substrate, on one hand, and a microscopic quantum field theoretic method to calculate the many-body effect, on the other, which changes the dielectric constant of the sample considerably even at the level of the one-loop contribution to the photon propagator for the finite or zero chemical potential of the fermions in the silicene.

We have considered two limits (see section 3) in this communication exhaustively. In the first limit, known as the long-wavelength limit, for the mass-less (and for the massive fermions) fermions we find that the plasmon resonance occur most definitely. We notice that the squared plasma frequency $\left(\frac{a\omega_{massive}}{v_F}\right)^2$ or $\left(\frac{a\omega_{massless}}{v_F}\right)^2$ for the massive as well as the massless fermions do not get influenced by the variation in the electric field $E_z$ at zero/ non-zero temperature. The quantity $\left(\frac{a\omega}{v_F}\right)$, however, depends on the strain field linearly as could be seen in Figure 5(c). The figure shows that the decrease in the tensile stain ($\varepsilon > 0$) and the increase in the absolute magnitude ($|\varepsilon|$) of the compressive strain($\varepsilon < 0$) brings about a decrease in the Plasmon frequency for the mass-less as well as the massive fermions in the long wavelength limit. We find that the valley-spin-locked (VSL) phase (represented by the upper curve in figure 5(c)) is characterized by the higher plasmon frequency for both the varieties of fermions compared to the topologically/trivially insulating phases represented by the lower curve in figure 5(c). This new piece of our finding is a clear-cut distinction between the phases and could be utilized to identify the finger-print of the onset of TPT. The TPT onset prediction through Casimir-Polder force (CPF) measurement is also possible. This has been hinted at in section3: We have plotted the dimensionless CPF given by Eqs. (15) and (16), at a given temperature and the sheet thickness, as a function of the ratio of the film thickness and the separation between the micro-particle and the film in Figure 2(b). The curve 1 corresponds to the case when the system is at TPT and the curve 2 when the system is at the topological/trivial insulating phase. We have mentioned in section 1 that the detection of TPT in silicene, by employing the Friedel oscillation and the collective excitation, had been reported earlier by Tabert et al.[22] and Chang et al[23]. We have also stated therein that one of our aims is to theoretically outline a novel method in addition to what had been presented by these authors. The novel result we mentioned about is that the magnitude of the force at a given ratio of the film thickness and the separation between the micro-particle and the film is greater at TPT than at the topological insulator and trivial insulator phases as could be observed from Figure 2(b). The reason being a finite chemical potential inhibits spatial dispersion and brings about an increase in the Casimir interaction substantially. However, with the emergence of a band gap the system becomes a poorer conductor and the dispersion effects start becoming more important[45,46]. Furthermore, it is easy to foresee that the optical conductivity[47] calculation at non-zero frequency may prove to be yet another reliable method to predict the onset of the topological transition. It must be pointed out that our investigation may also be dubbed as the one where the influence of the topological phase transitions on the Casimir-Polder force has been initiated for the first time. A similar investigation[48] in the context of the influence of Casimir energy on the critical field of a superconducting film had been reported several years ago. It was shown that it is possible to directly measure the variation of Casimir energy that accompanies the transition of the superconducting film. The problem in hand, however, has decidedly more complicated aspects, such as (i) the silicene layer is not flat, it is corrugated, and (ii) the retarded Casimir–Polder force [2,49] between two molecules is influenced by the presence of a third molecule due to the electromagnetic nature of this interaction. The part(ii) implies that the Casimir force is non-additive in that the force between two macroscopic bodies cannot be obtained by pair-wise summation of the interacting molecular interactions. Obviously, the effect of non-additivity has the potential to greatly complicate the predictions of the Casimir force for nonstandard geometries, such as corrugations. The investigation involving a corrugated silicene is a future task. The second limit is called the intrinsic limit. We find here that the collective charge excitations at zero doping, i.e., intrinsic plasmons, are absent in the pure Dirac case. This is shown in Figure 8 clearly in the $E_z - \varepsilon$ plane considered where real($\varepsilon(\omega, |\delta \boldsymbol{k}|, Ez, \varepsilon)) \neq 0$ everywhere. The plot refers to the moderate wavelength $((a|\delta \boldsymbol{q}|) \sim 0.1)$ and the low frequency $\left(\frac{a\omega}{v_F}\right) \ll 1$ case. The result-the absence of un-doped plasmons in the Dirac limit- above is not surprising as this is true for the in graphene as well as ordinary 2DEGs[50]. The (valley-spin-split) intrinsic plasmons, however, come into being in the case of the massive (gapped) Dirac particles with characteristic frequency close to 10 THz as shown in Figures (6) and (7).

The Casimir effect is attractive in most vacuum-separated metallic or dielectric geometries. Two electrically neutral spatially separated systems interacting via Casimir force will have access to the stable separation state only when the force transitions from repulsive at small separations to attractive at large separations. Such issues are important in the future development of micro- and nano-electromechanical systems (MEMS and NEMS). The most important advantage that the Casimir repulsion offers is in the operation of MEMS and NEMS, as it liberates one from the badgering problem of stiction. We have investigated and reported here the Casimir-Polder free energy corresponding to interactions of an electrically and magnetically polarizable[51]micro-particle with a magneto-dielectric sheet in section2. Our calculation pertains to the Faraday-Maxwell (static) limit, where the frequency dependence of all functions are ignored completely, resulting in the appearance of the conditions (15) and (16) above. It must be clarified that this limit is not the same as the low-temperature limit where $\omega_l = = 2\pi l k_B T/\hbar$ will get closer to each other and at zero temperature all of them contribute to dissipation. In order to look for the repulsive Casimir-Polder forces

**[52]** between the particle and film now, we assume that these objects may possess sheet-impedance ($Z_1$) and the polarizability ratio($Z_2$) values different from those of the natural materials, for example, silicene, graphene, etc.. We shall do some graphics now to see what does Eq.(16) convey for these hypothetical materials. Analyzing high temperature and the moderate temperature counterparts of (15) one may not gain probably a very different insight compared to what could be obtained from (15) directly. Therefore, the analysis of these results are not in the agenda at the moment. For the 2D graphics, we have assumed $\varepsilon_0^{(1)} = 14$, $Z_1 = 0.5$ and $1.00$, and $Z_2 \equiv \sqrt{(\eta(0)/\xi(0))} = (01, 02, 03, 04)$. We depict the crucial part of the Casimir-Polder interaction, viz. the function $F(d, \varepsilon_0^{(1)}, Z_1, Z_2, \omega = 0)$ in the static limit. We have approximated it by a quintic function in $\left(\frac{D}{d}\right)$ in (16). In figure 9 we have plotted this quintic as a function of $\left(\frac{D}{d}\right)$. In figure $9(a)$ ($\varepsilon_0^{(1)} = 14$, $Z_1 = 0.5$, and $Z_2 = (01, 02, 03, 04)$), we find that interaction is attractive as long as $\left(\frac{D}{d}\right) \lesssim 0.2$, or, $d \gtrsim 5D$. For $\left(\frac{D}{d}\right) > 0.2$ (or, $d < 5D$), the interaction is repulsive, while in figure $9(b)$ ($\varepsilon_0^{(1)} = 14$, $Z_1 = 1.0$, and $Z_2 = (01, 02, 03, 04)$), we find that interaction is attractive as long as $\left(\frac{D}{d}\right) \lesssim 0.1$, or, $d \gtrsim 10D$. For $\left(\frac{D}{d}\right) > 0.1$ (or, $d < 10D$), the interaction is repulsive. The results (repulsion at smaller separation $\left(\frac{d}{D}\right) \sim 1$ and the attraction at larger value) depicted in Figure 9 were expected as the Casimir-Polder/Casimir forces are very closely linked with the van der Waals' force. As regards the 3D graphics in Figure (10), we have taken here extreme non-trivial values of the polarizability ratio of the micro-particle($Z_2 = 20$) in order to examine whether the tale of non-natural materials have some ingredients where the strange and *bizarre* is blurred beyond clarity. A 'spot of vulnerability', for example, appears at $Z_1 < 1$ for $d \lesssim 4D$, where the repulsion suddenly changes to attraction followed by a swift comeback. The sheet used for obtaining Figure 10 may be termed as a moderate dielectric($\varepsilon_0^{(1)} = 14$). We, thus, notice from figure (9) that, for the repulsion to occur, the sheet material should be relatively high in the magnetic response (and the micro-particle has moderate magnetic polarizability). Generally, it is known **[53,54]** that for this purpose one requires a magnetic response strong enough to dominate the electric response of the material in a broad range of frequencies. Since this stringent condition is not met by any natural material, there has been a quest for an artificial material whose properties could be tailored in this direction. We have also investigated the Casimir-Polder interaction in the high-temperature limit( not discussed in this communication). In the high temperature (or, the large separation) regime, the long time behavior of the ubiquitous dissipation is dominated by an exponential decay with a time constant given by the first Matsubara frequency $\omega_1 = 2\pi k_B T/\hbar$. The attraction-repulsion crossover criterion is found to be a generalization of Eqs. (15) and (16).

**5. Conclusion** In conclusion, we note that the plasmonic resonances depicted in Figures 6 and 7 may occur for parameters suitable for the experiments like the Raman spectroscopy on 2+1 dimensional spin-orbit coupled (SOC) fermions on the low-buckled honey-comb structured lattice plane. The Plasmon mode undergo level broadening (or damping) as well, as manifested by the line width of the experimental plasmon peak. The level broadening agents which change the plasmon peak from a pure delta-function like pole in the response function to an approximate broadened Lorentzian shape are the ones arising from the (non-magnetic) impurity elastic scattering and the inherent Landau damping. In our calculation, Figure 7, in particular, indicates that the single-particle-excitation (SPE) continuum ($\text{Im}(\varepsilon(\omega, |\delta \boldsymbol{k}|, E_z, \varepsilon)) \neq 0$) also occur in some patches in the $\omega - |\delta \boldsymbol{k}|$ plane for the massive fermions. The decay of plasmons into e-h pairs, known as the Landau damping, occurs in the SPE continuum. Additionally, as the results of Iurov[44]et al. and our calculation have shown, these intrinsic Plasmon (associated with massive fermions) are a relatively strong and well-defined mode even at room temperatures (T~300K). This novelty of the mode definitely draws one's attention towards the mode's suitable candidacy for possible plasmonic applications.

The Casimir / Casimir-Polder effect is attractive in most vacuum-separated metallic or dielectric geometries. As has been already noted, two electrically neutral spatially separated systems interacting via Casimir force will have access to the stable separation state only when the force gets transformed from repulsion at small separations to attraction at large separations. We have investigated in this paper the Casimir-Polder free energy corresponding to interactions of a magnetically and electrically polarizable micro-particle with a silicene film and other non-natural material films as well. Our study shows that such an interaction is tunable in strength and sign. The latter, particularly, is true provided we go beyond the natural materials and look for the meta-materials fabricated at scales between the micron and the nanometer. In fact, our graphical representations reveal that a unusual magnetic response must be present over the electric response of the material under investigation in a broad range of frequencies for the Casimir-Polder repulsion to become a reality. On a quick side note, the frequency (ω) dependences of the permeability (μ) and the permittivity (ε) of the sheet material must be taken into consideration (these dependences are ignored here), for important information about the optical properties of the surface is encoded in these response functions. It is being hoped, notwithstanding the fact that the in-depth investigation of the present problem requires dealing with a quantum-mechanical description, that our semi-phenomeological approach with the Stoner-like criterion (of ferromagnetism) enshrined in Eqs.(15) and (16) for the attraction-repulsion crossover, will generate interest among the Casimir Physics community to search for new materials.

The prediction regarding artificial materials, such as the meta-materials[53,54] (MM) and the chiral meta-materials [55,56](CMM), with tunable magneto-dielectric properties had fuelled the hope of realizing the Casimir/CP repulsion and nano-levitation effect on demand in the second half of the last decade. The quest for the exotic material capable to deliver the Casimir/CP repulsion appeared to have been met with initial success. The hope, however, was dashed as the very conjecture of accessing the repulsive Casimir effect based on the CMMs was adjudged to be doubtful[57]. The reason shown by the authors [57] is that the proposal is irreconcilable with the causality and the passivity of the meta-materials. This had perhaps pushed the investigation trail back to the initial step. The recent developments in nano-fabrication/ design procedure of MMs [58,59] with specially tailored magneto-electric properties, however, have resulted in the regeneration of hope in the field on investigation of dispersion forces in the presence of MMs.

**Figure Caption**

**Figure 1:** The configuration of a micro-particle in vacuum characterized by the electric polarizability ξ(ω) and the magnetic polarizability η(ω) at a distance 'd' above a sample consisting of thin silicene film of thickness 'D' deposited on a thick substrate. While the film is characterized by the dielectric permittivity $\varepsilon^{(1)}(\omega)$ and the magnetic permeability $\mu^{(1)}(\omega)$, the substrate is by the permittivity $\varepsilon^{(2)}(\omega)$ and the permeability $\mu^{(2)}(\omega)$. We have chosen the coordinate plane (x, y) coinciding with the upper film surface and the z axis perpendicular to it.

**Figure 2:** (a)A plot of the dimensionless Casimir-Polder Force (CPF), at a given temperature and the sheet thickness, as a function of the ratio of the film thickness and the separation between the micro-particle and the film. We take the bare dielectric permittivity and the magnetic permeability of film material, respectively, to be $\varepsilon_0^{(1)}$ = 12 and $\mu_0^{(1)}$=1. At large momentum, the 'effective' dielectric constant of the sample is 19.1667. We find that the magnitude of the force and the free-energy decreases monotonically as the distance between the micro-particle and the sheet increases and vice versa.(b) A plot of the dimensionless Casimir-Polder Force (CPF), at a given temperature and the sheet thickness, as a function of the ratio of the film thickness and the separation between the micro-particle and the film when the system is at TPT(curve1) and when elsewhere(curve 2). We find that the magnitude of the force at a given ratio of the film thickness and the separation between the micro-particle and the film is greater at TPT than at the topological insulator and trivial insulator phases.

**Figure 3:** A 2D plot of the real and the imaginary parts of the static dielectric function as a function of the dimensionless wave number in the limit $a\delta q > 2ak^F_{\xi,s_z,\varepsilon}$. The curves show that the upper limit of the momentum integration in Eq.(25) may be fixed at 0.2, for the real and the imaginary parts tend to zero as $a\delta q \to 0.2$.

**Figure 4:** The vacuum polarization or charge screening in one loop approximation. The two internal lines corresponds to the two fermion Green's function and the wiggly lines to photons.  Similar diagrams could be considered for the other exchange particles (gauge bosons) in the appropriate context, such as for the gluons corresponding to the strong force between quarks.

**Figure 5. (a)** The plots of the dimensionless spectral gap $\Delta_{\varepsilon, E_z=0}$, as a function of the biaxial strain field ($\varepsilon$), and its least square fit. **(b)** A 3D-plot of the gap function $G_{\xi s_z}(E_z, \varepsilon)$, with the characteristic W-shape[16], as a function of $E_z$ and $\varepsilon$ for $\xi = +1$ and $s_z = -1$. This also corresponds to the phase diagram of the system in the $\varepsilon - E_z$ plane. We, further, find that for a given value of $\varepsilon$, $G_{\xi s_z}(E_z, \varepsilon)$ decreases to zero as $|E'_z|$ increases from zero to $|E'_{zc}|$ ($E'_{zc} = E'_{zc}(\varepsilon)$ is the critical field at which the gap function $G_{\xi s_z}(E_z, \varepsilon)$ becomes zero). The gap function increases, attains peak value, and decreases to zero as the electric field varies as $-E'_{zc} \leq E'_z \leq E'_{zc}$. **(c)** A plot of $(a\omega/v_F)$ for a given wave vector $(a|\delta q|) \ll 1$ (the Plasmon frequency in the long wavelength limit) as a function of $\varepsilon$ for $\xi = +1$ and $s_z = -1$.

**Figure 6.** Plots of $\varepsilon(\omega, |\delta k|)$ as function of $\omega$ in the limit $\mu' < \Delta_{\xi, s_z}$. It is evident from the plot (a) that the gapless (or massless) fermions do not yield the surface plasmon resonance. In the case (a), we have $E_z/E_c = 1$. The collective mode is obtained solving the equation real $(\varepsilon(\omega, |\delta k|)) = 0$ for the massive variety. We have assumed that the damping of the collective mode is weak. The two curves in each of (a), (b) and (c) correspond to the resonance modes for the two variety of spins. While $E_z/E_c < 1$ for (b), it is greater than one in (c). The plasmon frequency, by-and-large, is of order 10 THz..

**Figure 7.** (a) The contour plot of $\varepsilon(\omega, |\delta k|)$ for the mass-less fermions in the ($\delta k, \omega$) space in the intrinsic limit ($\mu' < \Delta_{\xi, s_z}$) with the electric field ratio $E_z/E_c = 1$. The surface Plasmon resonance (SPR) is absent in both the lower region and the upper region. (b) The yellow patches in the ($\delta k, \omega$) space correspond to the real part of the polarization function zero for the massive fermions when $E_z/E_c = 1$ and, therefore, they form the SPR regions for the critical electric field. (c) [(d)] The contour plot of $\text{Im}(\varepsilon(\omega, |\delta k|))$ as a function of ($\delta k, \omega$) for the mass-less **[massive]** fermions with $E_z/E_c = 1$. The SPE continuum ($\text{Im}(\varepsilon(\omega, |\delta k|)) \neq 0$) is everywhere except in the burnt-red patch where $\text{Im}(\varepsilon(\omega, |\delta k|)) = 0$. The decay of plasmons into e-h pairs, known as the Landau damping, occurs in the SPE continuum. (e)[(f)] The contour plot of $\text{Re}((\varepsilon(\omega, |\delta k|))$ as a function of ($\delta k, \omega$) for the massive fermions with $E_z/E_c = 0.5$ [$E_z/E_c = 1.5$]. In both the cases, there is possibility of the formation of the SPR patches.

**Figure 8.** The contour plot of $\varepsilon(\omega, |\delta k|, E_z, \varepsilon)$ in the intrinsic limit ($\mu' < \Delta_{\xi, s_z}$) for the mass-less fermions as function of $E_z$ and $\varepsilon$ (for $\xi = +1$ and $s_z = -1$).. An attempt has been made to solve the equation real $(\varepsilon(\omega, |\delta k|)) = 0$ for the mass-less fermions in the moderate wavelength $(a|\delta q|) \sim 0.1$ and the low frequency $\left(\frac{a\omega}{v_F}\right) \ll 1$ limit. As we do not get real $(\varepsilon(\omega, |\delta k|)) = 0$ it is obvious that the intrinsic Plasmon resonance is absent in the moderate wavelength- low frequency limit. The figure shows that the critical field $E'_{zc} = -\xi s_z \frac{\Delta_{soc}(\varepsilon)}{\ell(\varepsilon)}$, at which the gap function $G_{\xi s_z}(E_z, \varepsilon)$ becomes zero, is a function of the strain field : $E'_{zc} = E'_{zc}(\varepsilon)$. Thus, the strain field affects TPT to the extent of the shift in the value of the critical field. The 3D-plot of $\varepsilon(\omega, |\delta k|, E_z, \varepsilon)$ in the intrinsic limit ($\mu' < \Delta_{\xi, s_z}$) with the characteristic W-shape[16], in fact, corresponds to the phase diagram of the system in the $\varepsilon - E_z$ plane.

**Figure 9.** The 2D plots of the quintic $F(d, \varepsilon_0^{(1)}, Z_1, Z_2, \omega = 0)$ as a function of (D/d); the remaining parameters are held fixed. **(a)** Here we have taken $\varepsilon^{(1)} = 14$, $Z_1 = 0.5$, $Z_2 = $ (01, 02, 03, 04). The quintic function is positive, i.e. the Casimir-Polder interaction is attractive as long as (D/d) < 0.2, or, $d \gtrsim 5 D$. For (D/d) > 0.2 (or, d < 5D), the interaction is repulsive. **(b)** Here . $Z_1 = 1.00$. The Casimir-Polder interaction is attractive as long as (D/d) < 0.1.

**Figure 10.** A contour plot of the quintic F as a function of (D/d) and $Z_1$ is shown here. The dielectric constant of the sheet material $\varepsilon^{(1)} = 14$ (a moderate dielectric). The polarizability ratio $Z_2 = 20.00$. An 'Achilles' heel' (a spot of vulnerability) appears for d < 4D as $Z_1$ is increased from zero. At $Z_1 \approx 0.25$ the repulsion suddenly changes to attraction followed by a swift comeback.

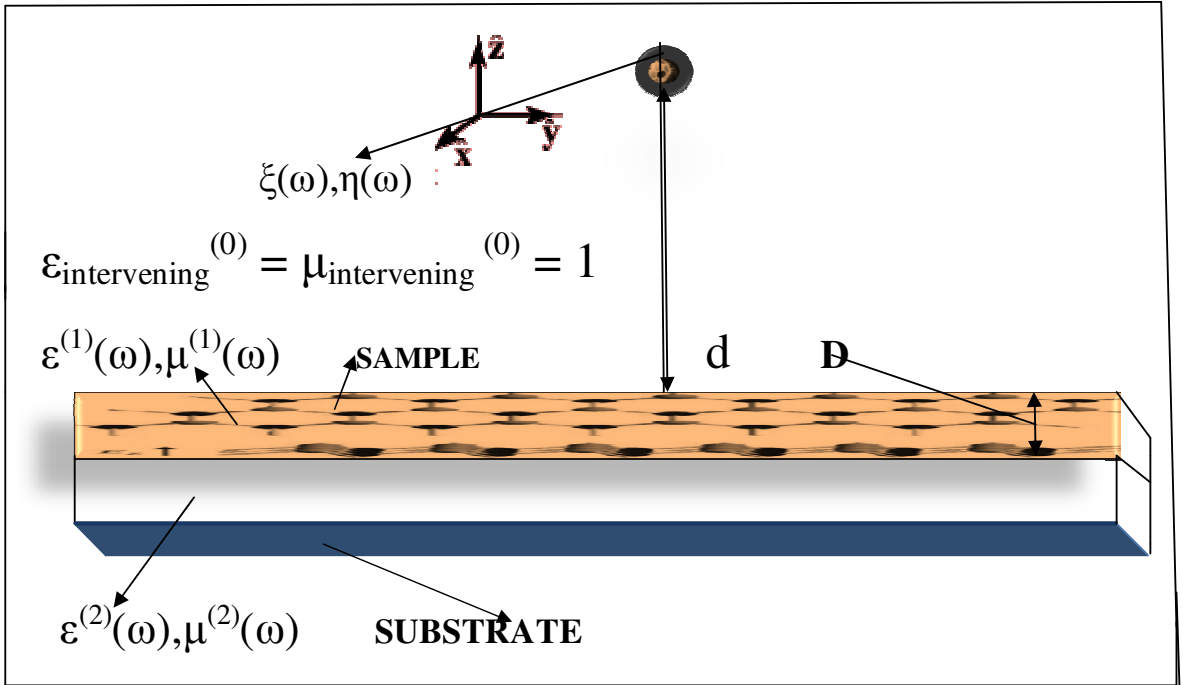
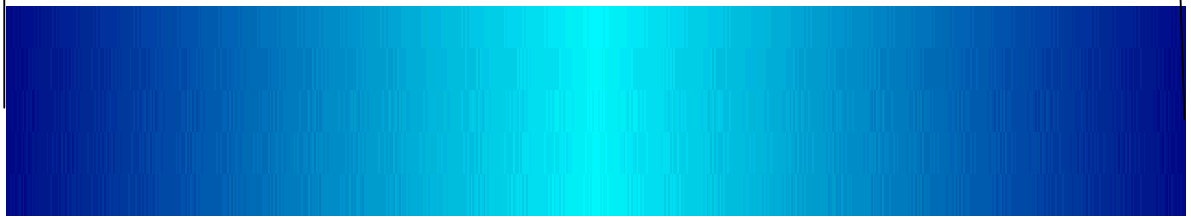

**FIGURE 1**

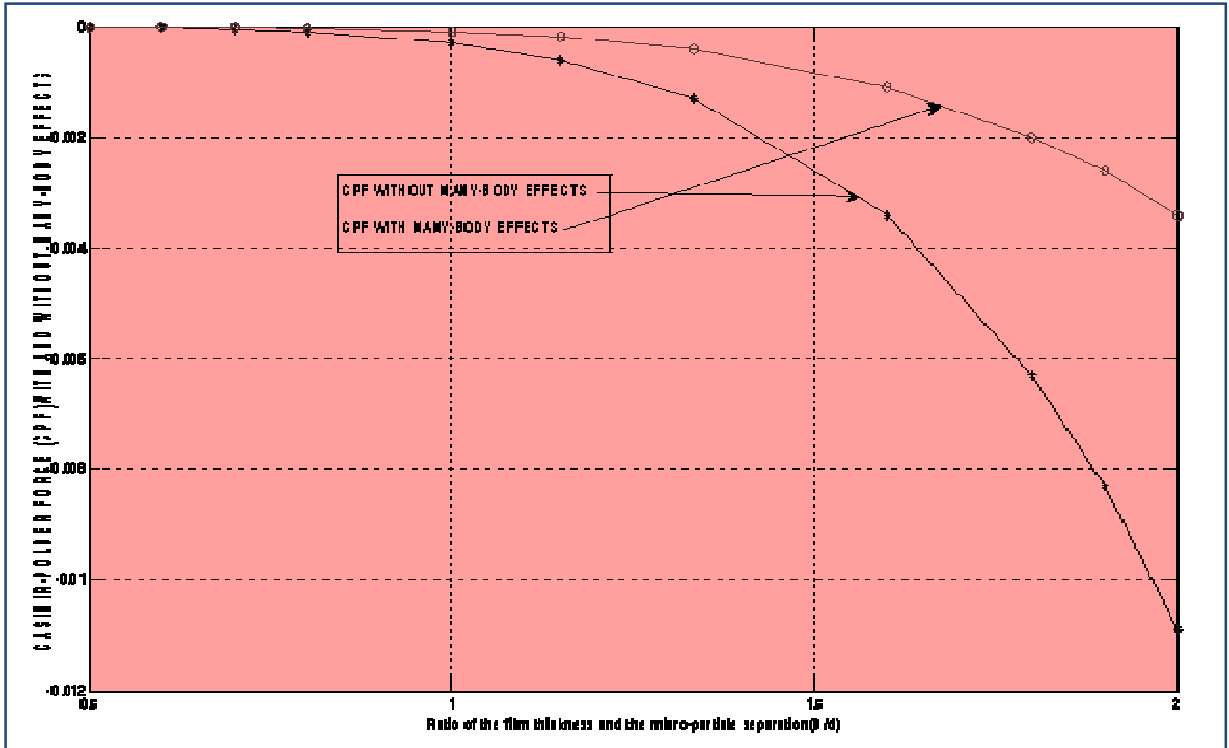

(a)

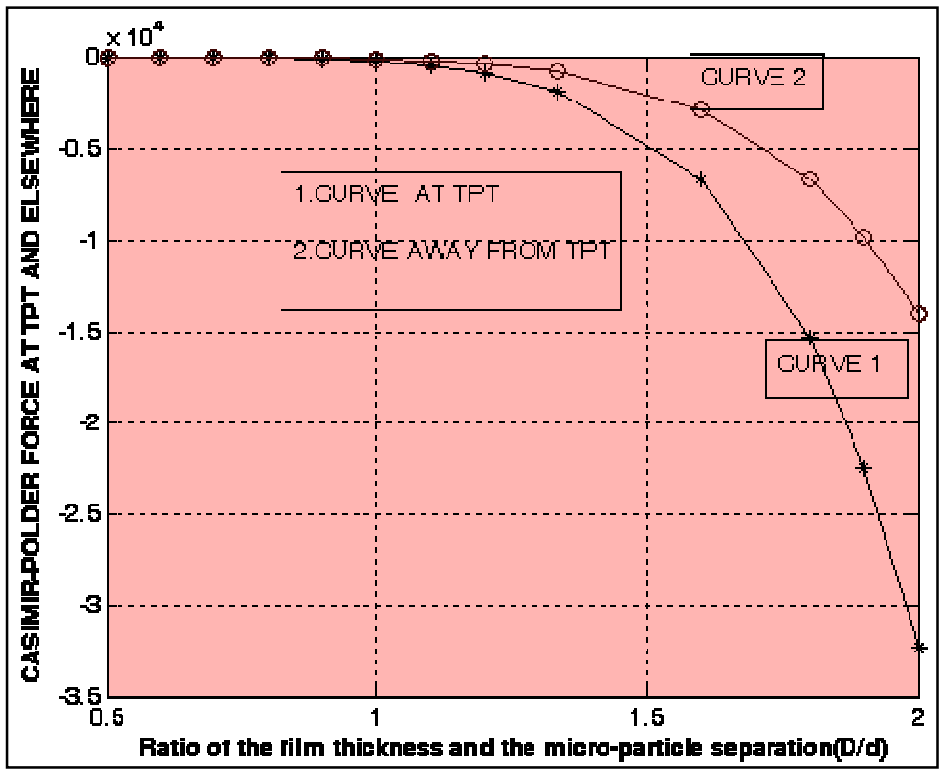

(b)

**FIGURE 2**

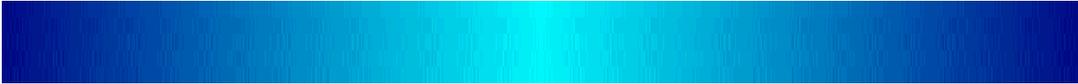

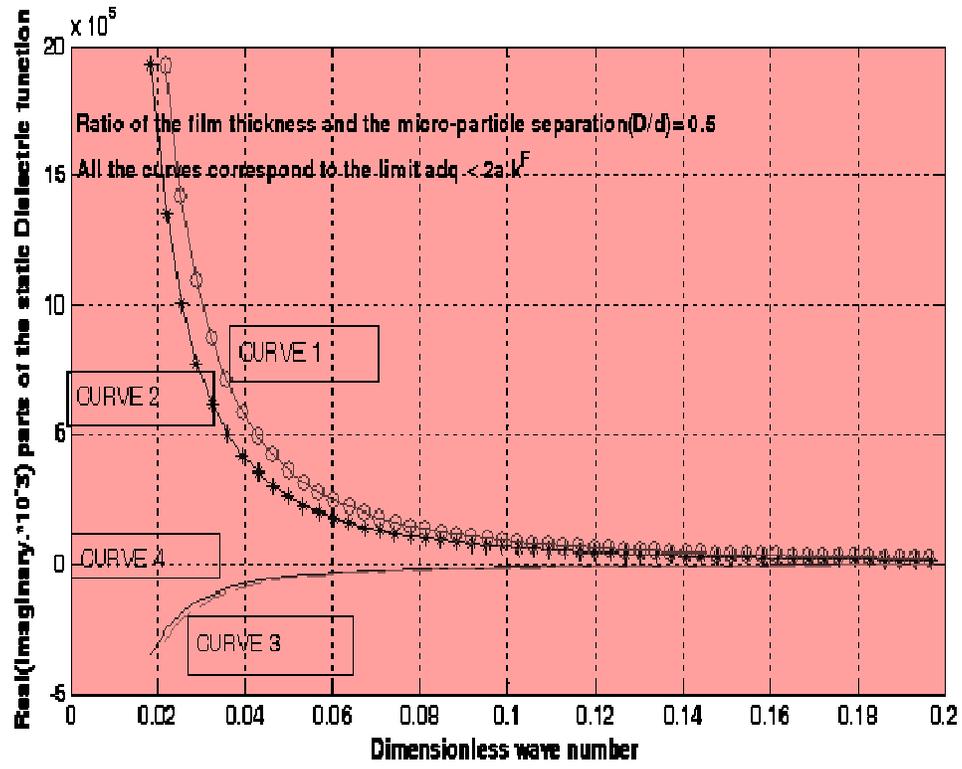

**FIGURE 3**

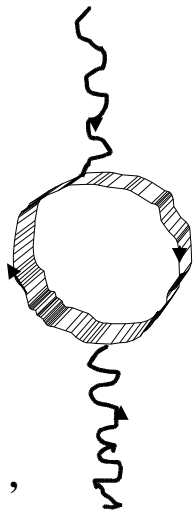

$\delta k+$

,

**FIGURE 4**

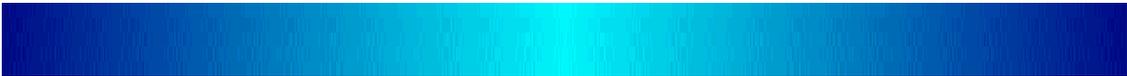

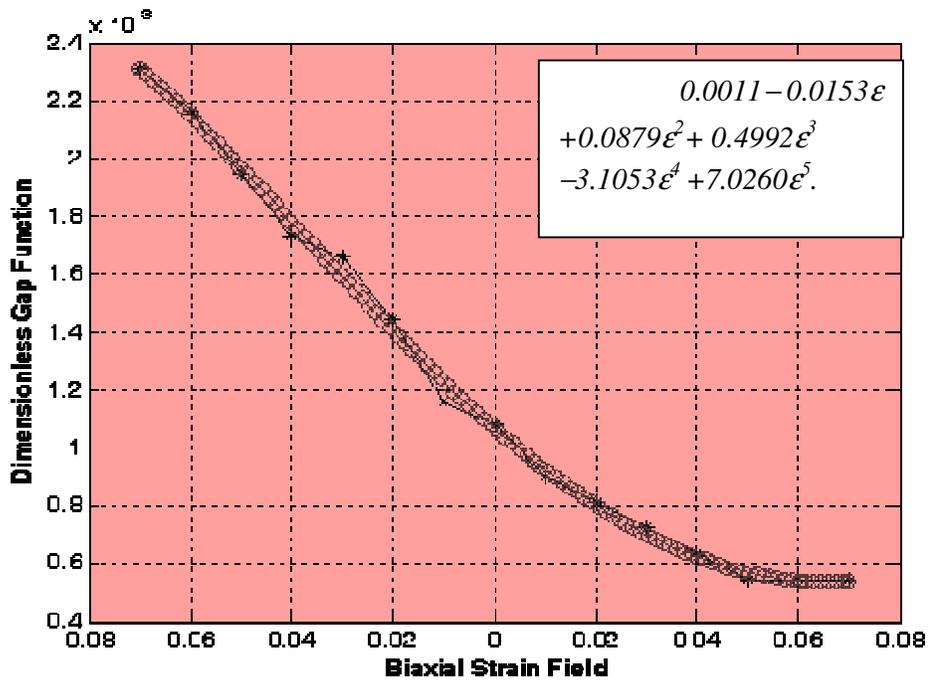

$0.0011 - 0.0153\varepsilon$
$+0.0879\varepsilon^2 + 0.4992\varepsilon^3$
$-3.1053\varepsilon^4 + 7.0260\varepsilon^5.$

(a)

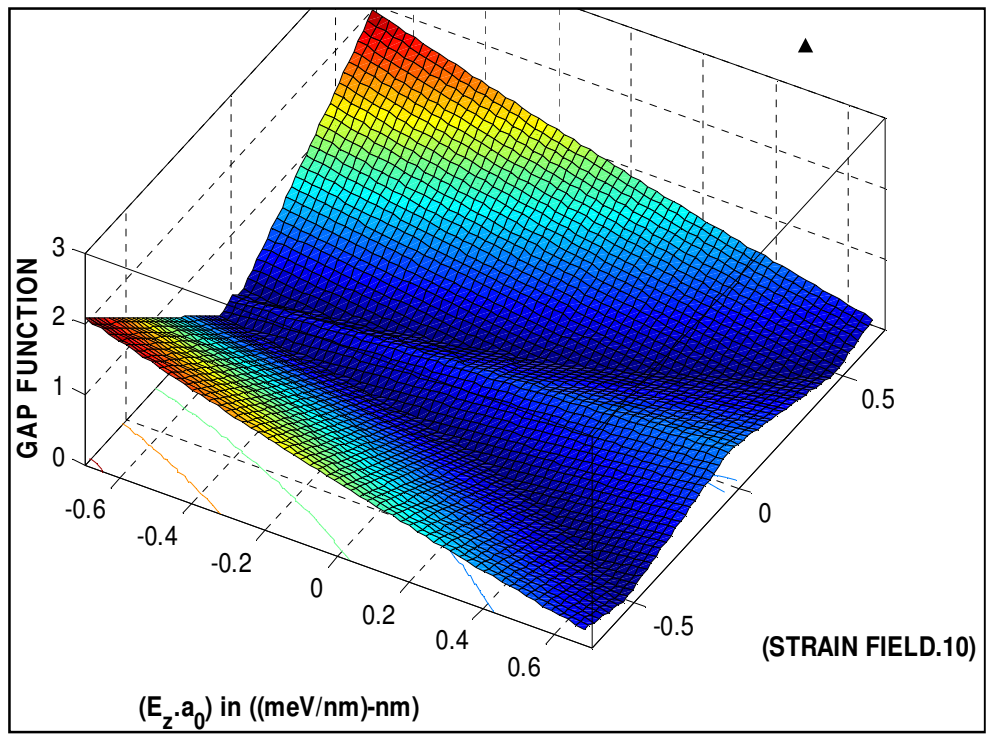

(b)

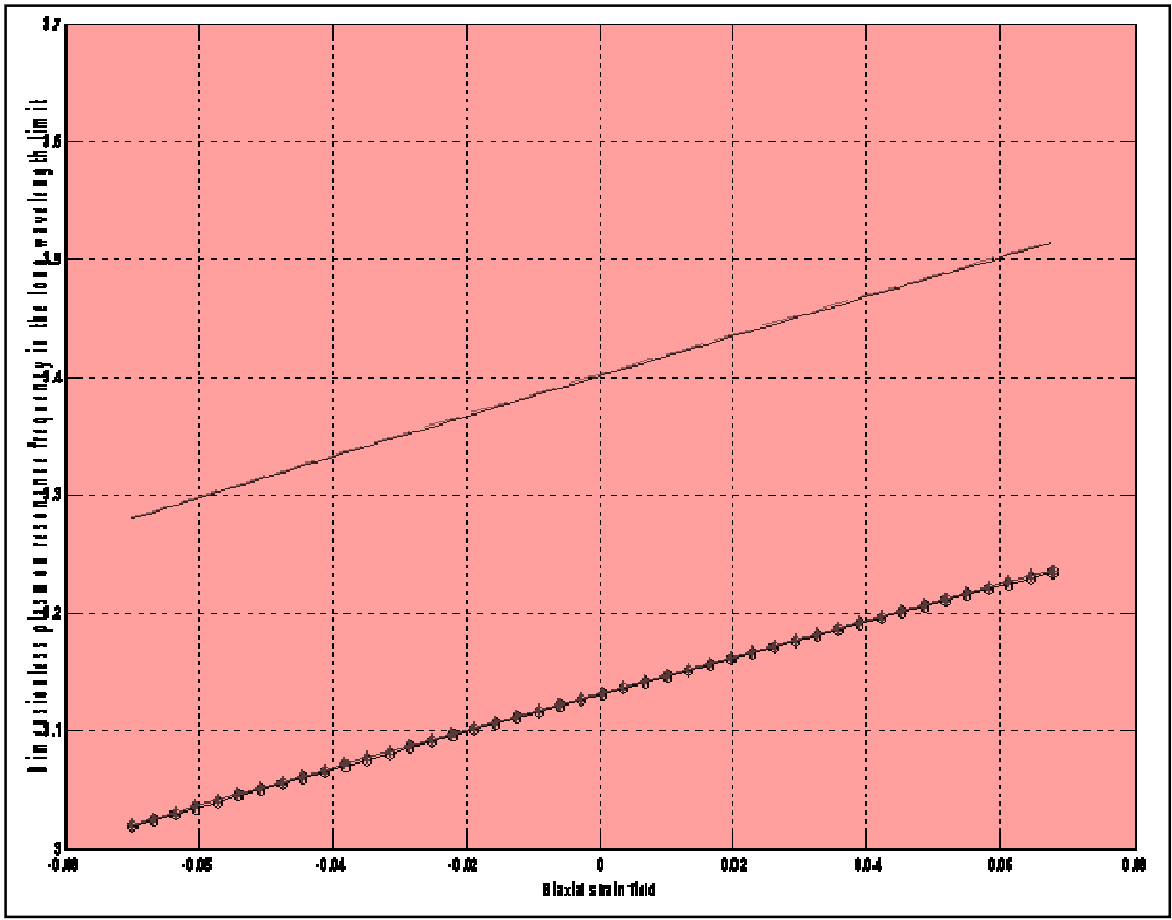

(c)

**FIGURE 5**

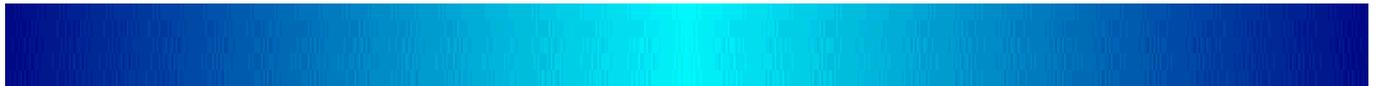

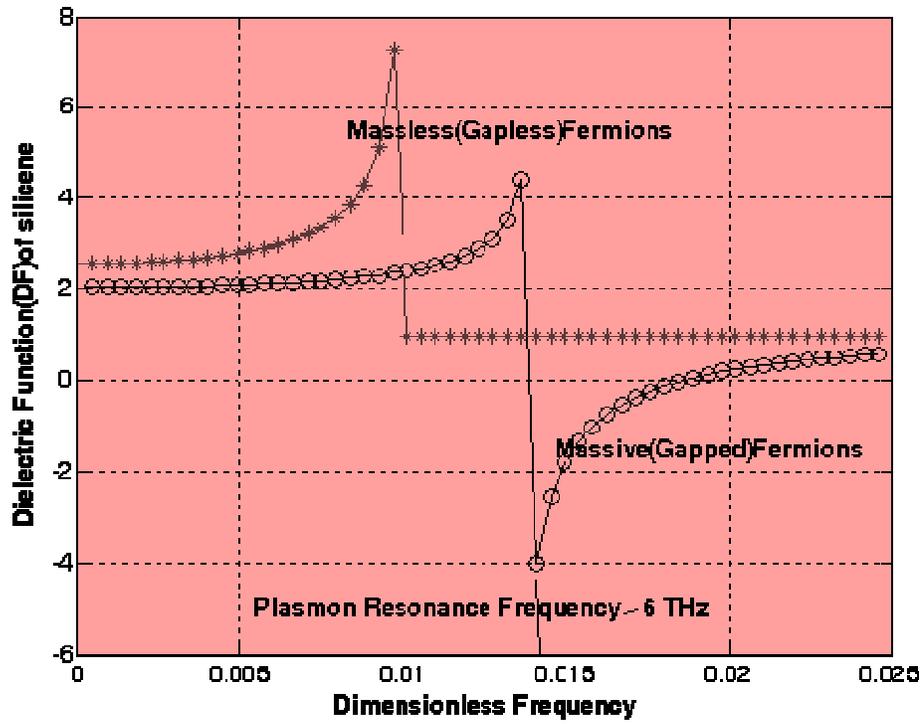

(a)

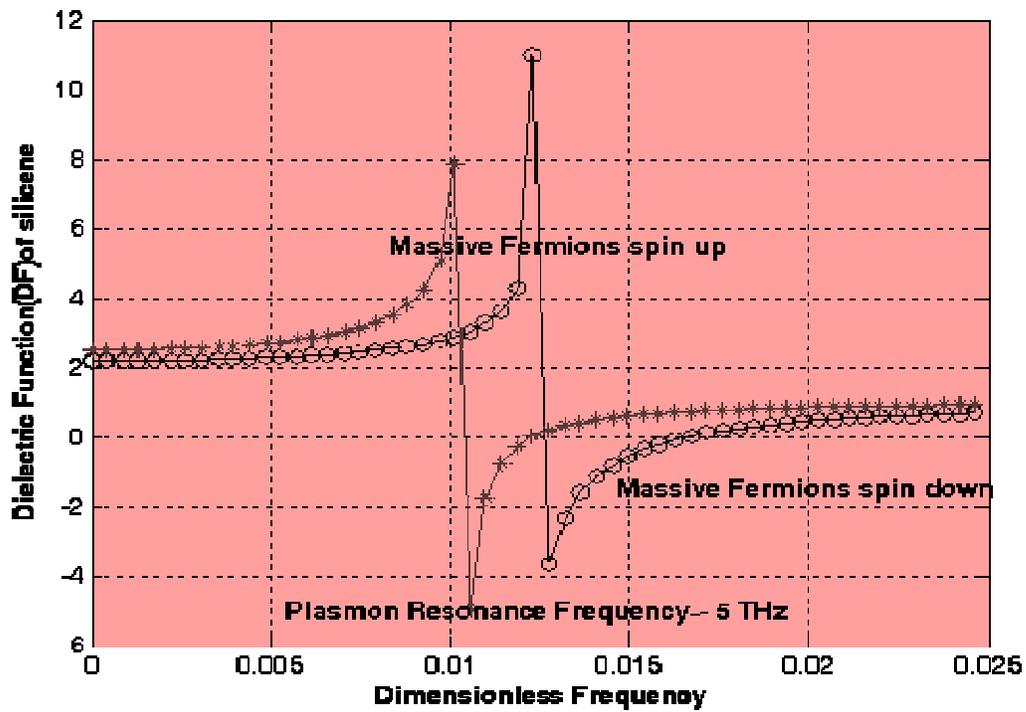

(b)

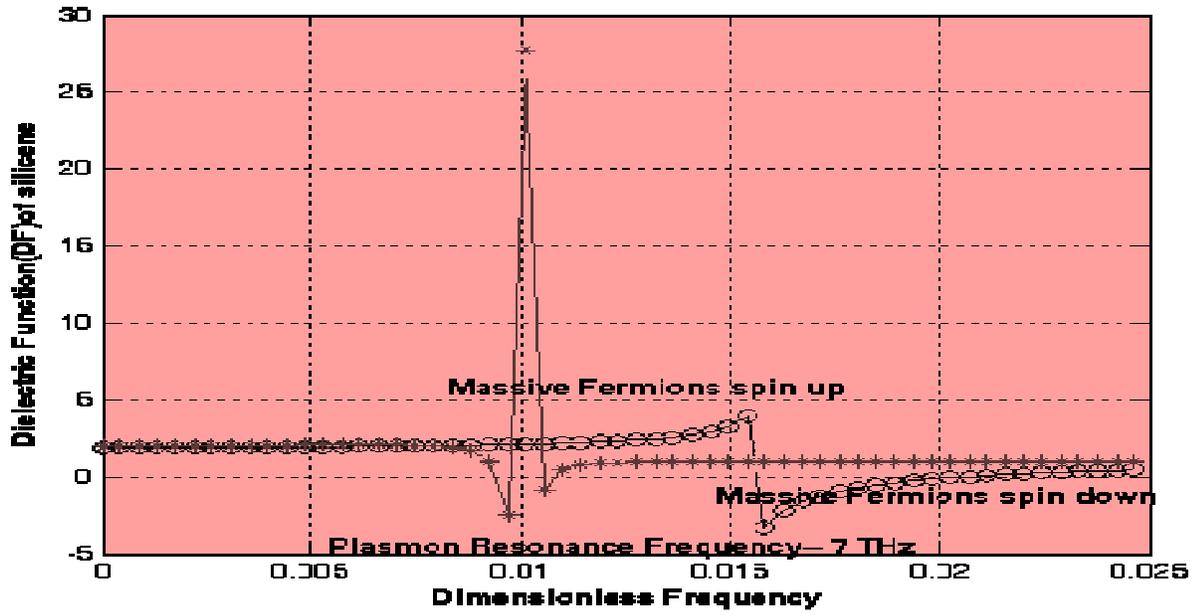

FIGURE6

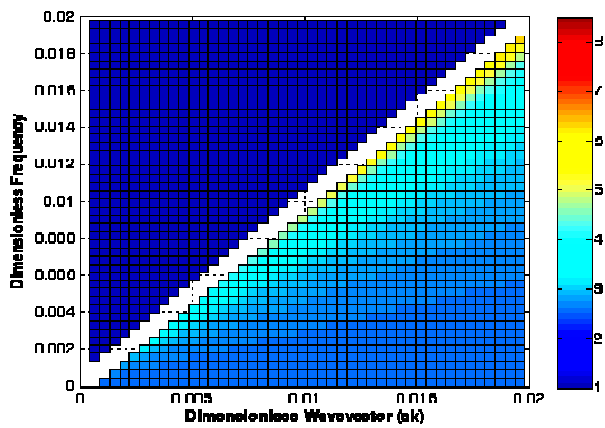
(a)

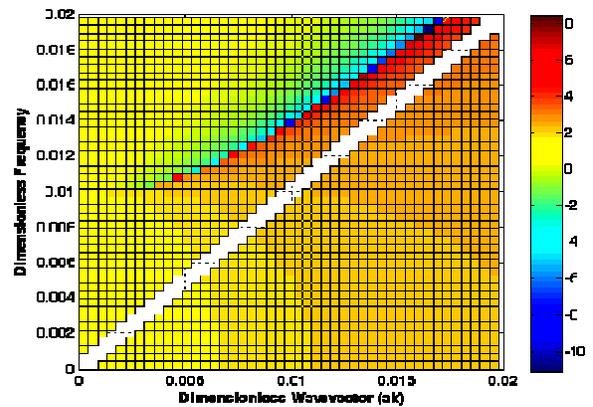
(b)

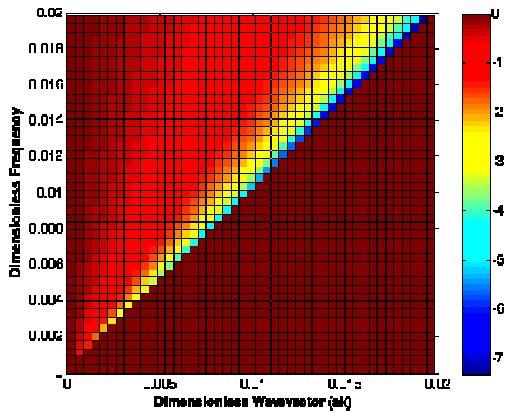

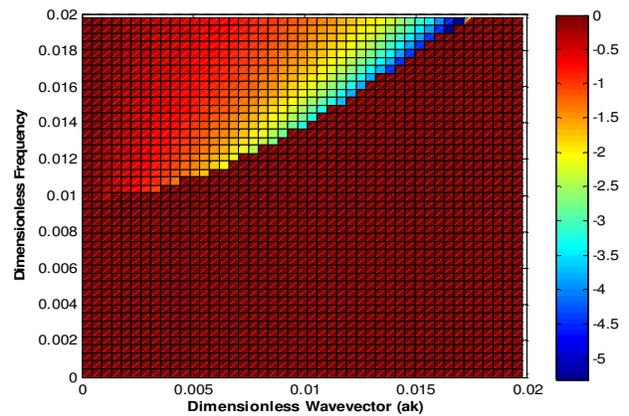

**(c)** **(d)**

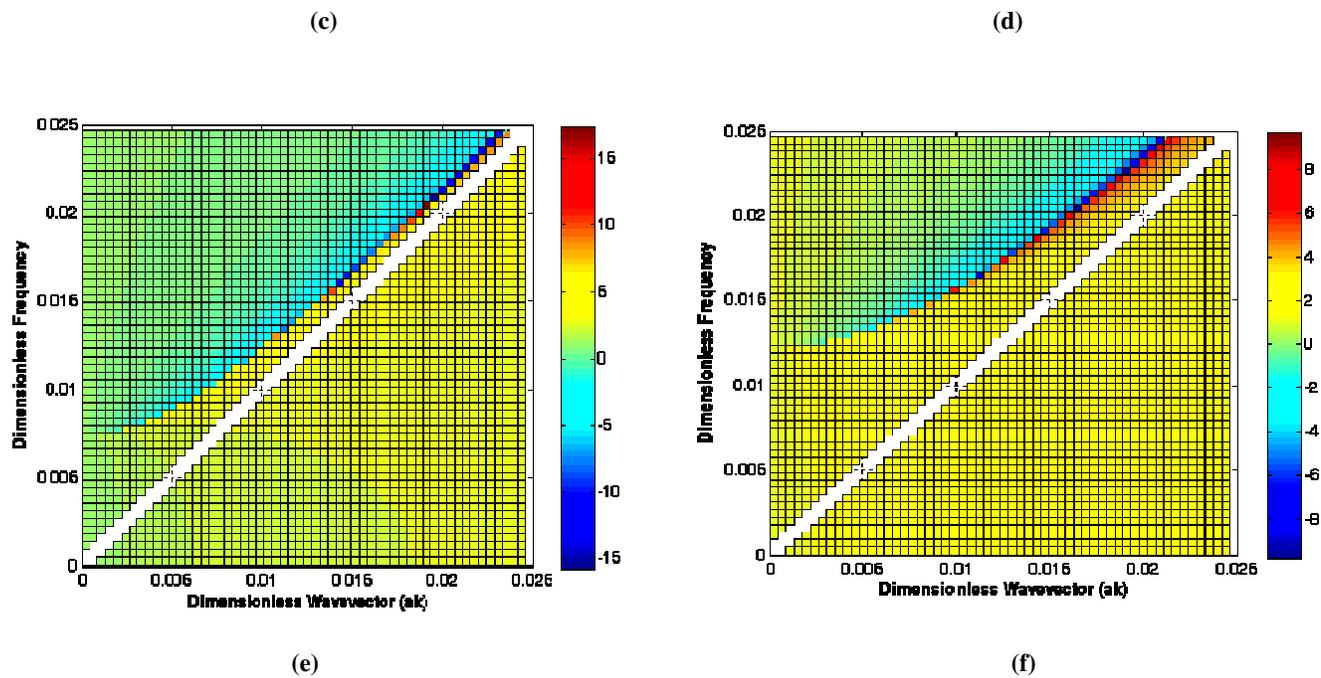

**(e)** **(f)**

**FIGURE 7**

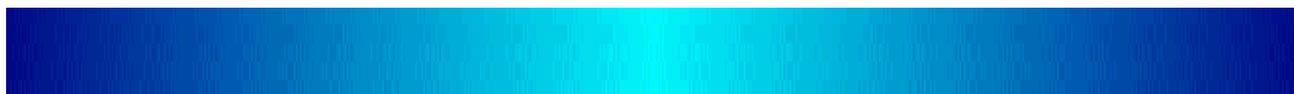

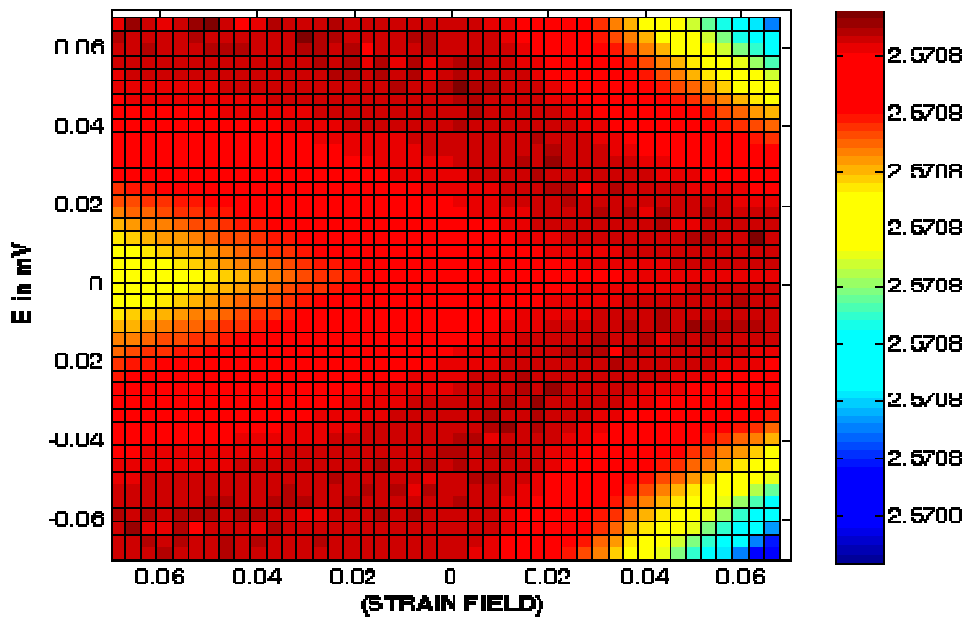

**FIGURE 8**

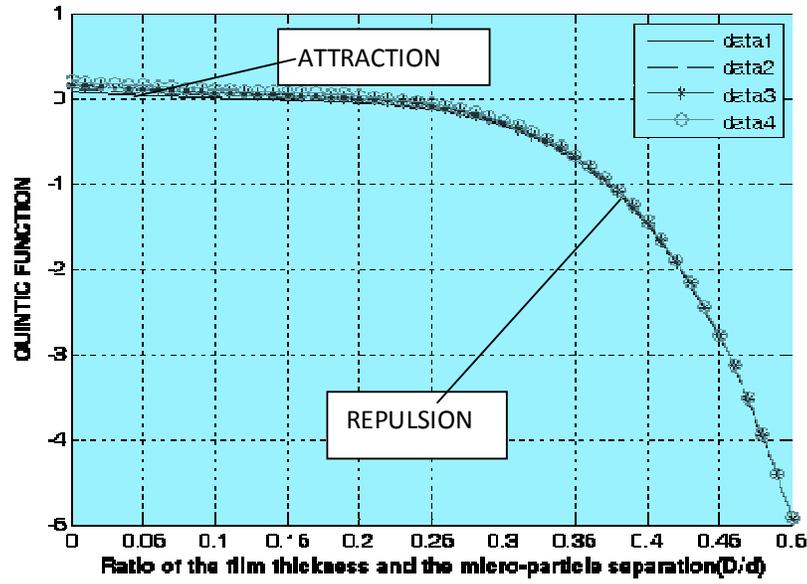

(a)

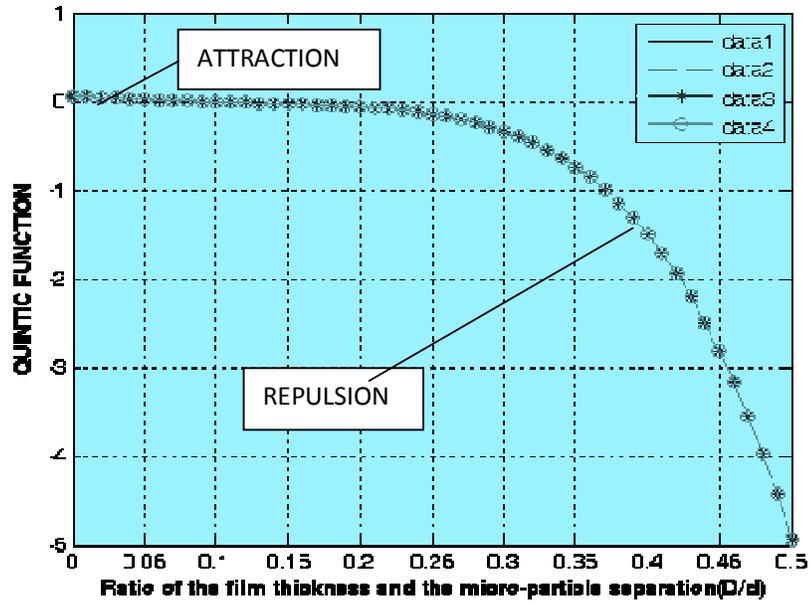

(b)

**FIGURE 9**

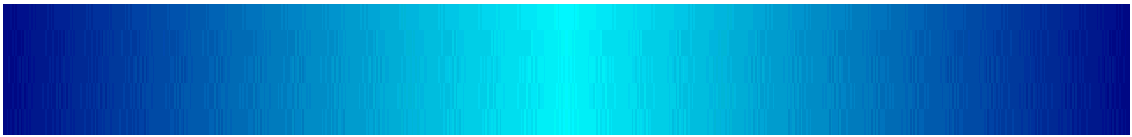

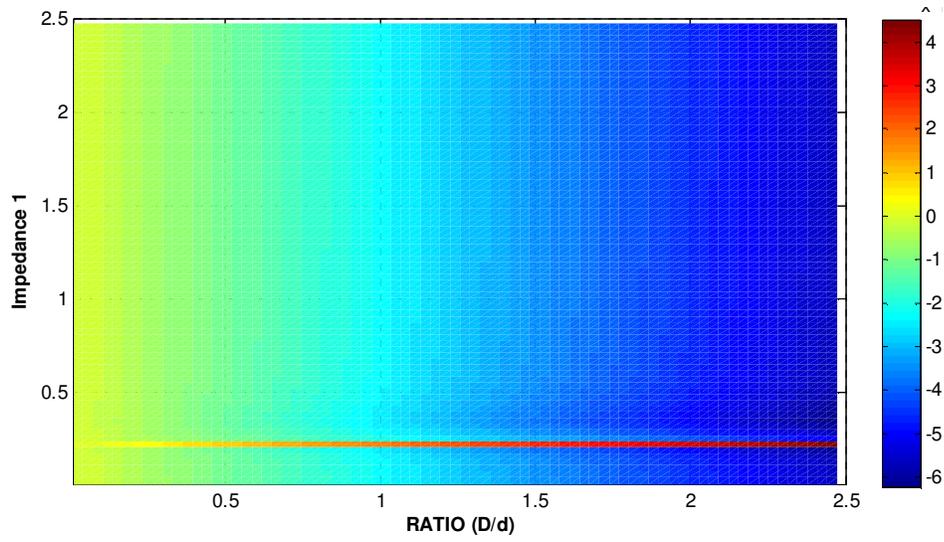

**FIGURE 10**

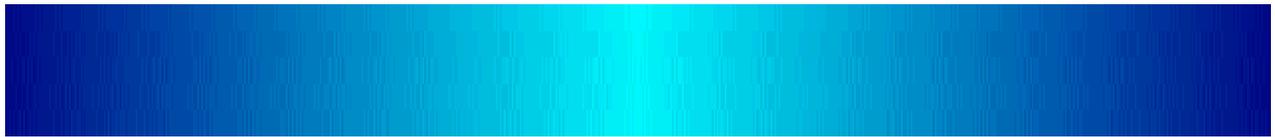

## Table and Legend

**Table 1.** The spin-valley locking of mass-less and massive fermions.

| Valley / Spin | K<br>Iso-spin<br>= +1 | K′<br>Iso-spin<br>= −1 |
|---|---|---|
| Spin-down State(↓) | Gap-less<br>(Mass-less) | Gapped<br>(Massive) |
| Spin-up State (↑) | Gapped<br>(Massive) | Gap-less<br>(Mass-less) |